\documentclass{article} % For LaTeX2e
\usepackage{iclr2027_conference,times}

\usepackage{amsmath,amsfonts,bm}

\def\eqref#1{equation~\ref{#1}}
\def\1{\bm{1}}

\DeclareMathAlphabet{\mathsfit}{\encodingdefault}{\sfdefault}{m}{sl}
\SetMathAlphabet{\mathsfit}{bold}{\encodingdefault}{\sfdefault}{bx}{n}

\usepackage[utf8]{inputenc} % allow utf-8 input
\usepackage[T1]{fontenc}    % use 8-bit T1 fonts
\usepackage{hyperref}       % hyperlinks
\usepackage{url}            % simple URL typesetting      % professional-quality tables
\usepackage{amsfonts}       % blackboard math symbols
\usepackage{nicefrac}       % compact symbols for 1/2, etc.
\usepackage{enumitem}
\usepackage{microtype}      % microtypography
\usepackage{xcolor}         % colors

\usepackage{booktabs} % For formal tables
\usepackage{makecell}
\usepackage{textcomp}
\usepackage{amsmath, amsthm}

\usepackage[ruled, vlined, linesnumbered]{algorithm2e}
\usepackage[noend]{algpseudocode}
\usepackage{pdfpages}
\usepackage{amssymb}
\usepackage{mathrsfs}
\usepackage{epsfig}
\usepackage{multirow}

\usepackage{balance}
\usepackage{wrapfig}
\usepackage[normalem]{ulem}
\usepackage{threeparttable}
\newcolumntype{C}[1]{>{\centering\arraybackslash}p{#1}}
\usepackage{lineno}

\usepackage{bbm}

\usepackage[table]{xcolor}
\usepackage{graphicx}
\usepackage{float}
\usepackage{subfig}

\definecolor{sftbg}{HTML}{FBECC8}    % yellow
\definecolor{dpobg}{HTML}{E6ECF7}      % blue
\definecolor{reasonbg}{HTML}{EEDFD1}   % brown
\definecolor{retrbg}{HTML}{E8E8E8}     % gray
\definecolor{rrcmbg}{HTML}{E8F3E5}     % green
\definecolor{ppink}{HTML}{FCE4EC}   % pink

\usepackage[most]{tcolorbox}

\definecolor{tagblue}{RGB}{0,0,220}
\definecolor{tagcyan}{RGB}{0,150,200}
\definecolor{tagbrown}{RGB}{160,100,40}
\definecolor{tagpink}{RGB}{200,0,100}

\newcommand{\thinkopen}{\textcolor{tagblue}{\texttt{\textless think\textgreater}}}
\newcommand{\thinkclose}{\textcolor{tagblue}{\texttt{\textless/think\textgreater}}}
\newcommand{\toolopen}{\textcolor{tagcyan}{\texttt{\textless tool\_call\textgreater}}}
\newcommand{\toolclose}{\textcolor{tagcyan}{\texttt{\textless/tool\_call\textgreater}}}
\newcommand{\respopen}{\textcolor{tagbrown}{\texttt{\textless tool\_response\textgreater}}}
\newcommand{\respclose}{\textcolor{tagbrown}{\texttt{\textless/tool\_response\textgreater}}}
\newcommand{\answeropen}{\textcolor{tagpink}{\texttt{\textless answer\textgreater}}}
\newcommand{\answerclose}{\textcolor{tagpink}{\texttt{\textless/answer\textgreater}}}

 \title{RRCM: Ranking-Driven Retrieval over Collaborative and Meta Memories for LLM Recommendation}

\author{
   Shijun Li$^{1}$,
   Pranav Belligundu$^{1}$,
   Tianxin Wei$^{4}$,
   Wooseong Yang$^{2}$,
   Yu Wang$^{3}$, 
   Joydeep Ghosh$^{5}$ 
   \\
  $^{1,5}$The University of Texas at Austin, $^{2}$University of Illinois at Chicago\\
  $^{3}$Capital One AI Foundations, $^{4}$University of Illinois Urbana-Champaign \\
  \texttt{shijunli@utexas.edu}\\
  \ \ \
  }

\iclrfinalcopy % Uncomment for camera-ready version, but NOT for submission.
\begin{document}

\maketitle
 \thispagestyle{fancy}
 \fancyhead[L]{Preprint. Under review.} % 或留空 \fancyhead[L]{}

\begin{abstract}
Large Language Models (LLMs) have emerged as a promising paradigm for next-generation recommender systems, offering strong semantic understanding and natural-language reasoning abilities. Despite recent progress, current LLM-based recommenders still face key challenges in constructing decision-relevant contexts from heterogeneous evidence. First, existing methods often rely on fixed context construction strategies: collaborative behavioral evidence and item-side metadata are typically incorporated through predefined prompts, static retrieval pipelines, or handcrafted injection mechanisms, making it difficult to determine what information is truly beneficial for each instance. Second, heterogeneous evidence introduces a severe context-efficiency bottleneck. Rich metadata and collaborative interaction records can quickly overwhelm the context window, while aggressive compression or heuristic filtering may discard fine-grained evidence critical for accurate recommendation.  To address these challenges, we propose RRCM, a ranking-driven retrieval-and-reasoning framework over collaborative and metadata memories for LLM-based agentic recommendation. RRCM starts from a lightweight user-history context and learns whether to recommend directly, retrieve collaborative evidence, retrieve item metadata, or interleave both through reasoning. Both memories are represented in natural language and accessed through a unified retrieval interface, enabling flexible evidence acquisition without handcrafted CF injection or fixed retrieval rules. We optimize this memory-reading policy with an outcome-only ranking reward, instantiated using group relative policy optimization, so that retrieval decisions are directly driven by final top-$k$ recommendation quality. Extensive experiments show that RRCM significantly outperforms traditional baselines and diverse LLM-based recommendation approaches. Code is available at: \href{https://anonymous.4open.science/r/RRCM-6233}{https://anonymous.4open.science/r/RRCM-6233}.
\end{abstract}

\section{Introduction}\label{sec:intro}

Large Language Models (LLMs) have recently emerged as a promising foundation for recommender
systems, not only as sequence models over user interaction histories but also as reasoning engines
that infer user preferences and support decision-making
\citep{wu2024survey,fang2025reason4rec}. Compared with traditional ID-based models, LLMs can leverage semantic knowledge, understand recommendation context, and generate interpretable rationales. However, recommendation remains a ranking problem driven by platform-specific evidence.
Effective LLM-based recommendation therefore requires constructing a decision-relevant context from heterogeneous sources, rather than simply relying on the LLM's parametric knowledge or placing all available information into the prompt.

Despite recent progress, two key challenges limit current LLM recommenders. First, existing methods often rely on fixed context construction strategies for heterogeneous recommendation evidence.
Collaborative behavioral evidence, such as cross-user and cross-item interaction patterns, is central to classical recommender systems \citep{mf,fm,kang2018self}, while item-side metadata provides important semantic grounding for understanding item attributes. Recent studies incorporate collaborative information through ID-like tokens, templated neighborhood summaries, or external collaborative embeddings \citep{zhang2024text,zhang2025collm}; other methods use predefined metadata retrieval or compression pipelines to enrich item representations
\citep{chen2024hllm,yi2025recgpt}. However, these approaches typically determine the evidence source and context format in advance. As a result, they do not explicitly learn what evidence is useful per instance, when it should be retrieved, or how it should be incorporated into
LLM's decision.

Second, heterogeneous recommendation evidence introduces a severe context-length and efficiency
bottleneck. Real-world item metadata, user histories, and collaborative interaction records can be
large and diverse. Directly including rich metadata for every item or many collaborative histories in
the prompt can produce prohibitively long inputs for training and low-latency inference
\citep{wu2024survey}. Existing remedies, such as heuristic filtering, compression, or hierarchical
preprocessing \citep{chen2024hllm,yi2025recgpt}, reduce token usage but may discard fine-grained evidence that is important for ranking. This issue is especially acute for niche, newly introduced, or ambiguous items, where title-only contexts can be insufficient and the LLM may hallucinate item properties when reasoning without grounded evidence \citep{longtailhallucination}. Thus, the central question is not how to provide more context by default, but how to adaptively construct the right context for each case.

To address these challenges, we propose \textbf{RRCM}, a {R}anking-driven {R}etrieval framework over
{C}ollaborative and item {M}emories for LLM-based agentic recommendation. RRCM treats
recommendation as an adaptive evidence acquisition and reasoning process. It starts from a
lightweight user-history context consisting primarily of item titles, which is often sufficient for
popular items or clear sequential patterns. When additional evidence is needed, the model can query
a unified retrieval corpus containing two complementary memories: {collaborative memory}, which
stores anonymized interaction records from historical users, and {meta memory}, which stores rich
item metadata. Both memories are represented in natural language and accessed through a single
retrieval interface, allowing the model to flexibly acquire behavioral evidence and item-side features
without handcrafted CF injection or fixed metadata retrieval rules.

% RRCM moves beyond fixed retrieval rules and handcrafted CF-injection mechanisms by learning a policy that decides when retrieval is necessary, which memory to retrieve from, and how to formulate natural-language queries. Retrieval and reasoning can be interleaved across multiple steps: the model may retrieve item-side metadata to disambiguate a niche item, retrieve collaborative histories to identify behaviorally supported candidates, and finally recommend an item grounded in both sources of evidence. Since the optimal retrieval timing, evidence type, and query formulation are unknown, we optimize this policy through outcome-driven reinforcement learning. The training reward is defined using top-$k$ ranking metrics together with a format-validity term, and we instantiate this ranking-driven objective with Group Relative Policy Optimization \citep{grpo}, encouraging the model to retrieve selectively only when external evidence can improve final recommendation quality. 

RRCM then learns a policy that decides when retrieval is necessary, which memory to retrieve from, and how to formulate natural-language queries. Retrieval and reasoning can be interleaved across
multiple steps: the model may retrieve item-side metadata to disambiguate a niche item, retrieve
collaborative histories to identify behaviorally supported candidates, and finally recommend an item
grounded in both sources of evidence. Since the optimal retrieval timing, evidence type, and query
formulation are unknown, we optimize this policy through outcome-driven reinforcement learning.
The training reward is defined using top-$k$ ranking metrics together with a format-validity term, and
we instantiate this ranking-driven objective with Group Relative Policy Optimization \citep{grpo},
encouraging the model to retrieve selectively only when external evidence can improve final
recommendation quality. Our main contributions are summarized as follows:

\begin{itemize}[leftmargin=*]
    \item \textbf{Unified heterogeneous memory retrieval.}
    We formulate LLM-based recommendation as a ranking-driven context construction problem over
    heterogeneous recommender memories, and propose a dual-memory retrieval framework that
    represents both anonymized collaborative histories and rich item metadata in natural language.

    \item \textbf{RL-based adaptive retrieval and reasoning.}
    We introduce RRCM, which learns when to retrieve, which memory to retrieve from, and how to
    reason over retrieved evidence. The policy is optimized with an outcome-driven ranking objective
    that directly uses recommendation quality as supervision.

    \item \textbf{Significant improvement with selective evidence acquisition.}
    Experiments show that RRCM consistently improves recommendation performance over strong
    traditional and LLM-based baselines, while learning to selectively retrieve collaborative/meta
    memories only when needed, reducing unnecessary context construction and improving efficiency.
\end{itemize}

\section{Related Work}\label{sec:related}

\noindent \textbf{LLMs for recommendation and reasoning.}
Recent work explores using LLMs as recommender systems by framing recommendation as language
modeling or conditional generation over user interaction histories and item descriptions \citep{p5,wu2024survey}.
Compared with conventional methods, LLM-based approaches can ingest richer
natural-language signals and can generate explanations or preference
summaries alongside predictions. In parallel, reasoning-enhanced recommendation can prompt or train
LLMs to produce intermediate rationales to improve both
accuracy and interpretability \citep{fang2025reason4rec,cotrec}. While these methods demonstrate that LLMs can go beyond pure pattern matching, they often still rely predominantly on the target user’s history and
candidate text, leaving broader collaborative signals underutilized and offering limited control over
when additional metadata should be consulted.

\noindent \textbf{Incorporating collaborative signals into LLMs.}
Collaborative filtering is the core ingredient in classical recommendation due to its ability to
exploit cross-user and cross-item structure \citep{mf,fm}.
Incorporating CF information into LLM-based recommendation has attracted increasing interest.
Representative directions include (i) augmenting the LLM interface with ID-like tokens or templated
neighborhood summaries \citep{zhang2024text}, and (ii) injecting externally trained collaborative embeddings as additional
inputs \citep{zhang2025collm}. Another pattern in practice is hybridization, where CF-style
candidate generation is followed by LLM re-ranking or explanation over a narrowed set of items
\citep{zhou2025hymirec}. However, most existing approaches use fixed integration schemes and do not
explicitly learn {when} collaborative evidence is needed under a limited context budget, and {how} to retrieve the most valuable collaborative information dynamically according to the current recommendation context.
RRCM instead treats collaborative evidence as retrievable textual memory within a unified corpus, enabling on-demand access through natural-language queries.

\noindent \textbf{Retrieval augmentation and outside information acquisition.}
Retrieval-augmented generation grounds LLM outputs by selecting relevant documents and is a standard approach to improve factuality and manage context length \citep{rag}. In recommendation, retrieval can fetch metadata snippets, reviews, behavioral statistics, or knowledge-graph facts, and has been used to enrich prompts and to mitigate long-tail hallucinations \citep{wang2025knowledge, kgrec}.
Nonetheless, these pipelines typically treat retrieval as a separate heuristic stage (retrieve top-$k$, then generate), which can retrieve unnecessarily and fail to capture collaborative regularities. More broadly, LLM agent work shows that models can interleave reasoning with external actions such as search or tool calls, and that learning {stopping} and {query formulation} is critical for efficiency \citep{react,selfrag,toolformer, jin2025search}. RRCM brings these ideas to recommendation by integrating retrieval into the decision process and optimizing the full loop end-to-end with outcome-based RL, encouraging efficient, evidence-driven recommendations. We've included a more detailed discussion of recent related work in Appendix \ref{sec:exp_dis}.

\section{METHODOLOGY}\label{sec:method}

We propose \textsc{RRCM}, a generative recommendation framework that formulates LLM-based recommendation as a ranking-driven context construction and reasoning process over heterogeneous memories. Instead of placing all available information into the input context or relying on fixed retrieval rules, \textsc{RRCM} learns to dynamically interleave internal reasoning with external memory access. % Specifically, the model decides whether to rely on its parametric knowledge, retrieve collaborative behavioral evidence, retrieve rich item-side features, or combine both types of evidence for the final ranking decision.

\subsection{Problem Formulation: Ranking-Driven Context Construction}

Let $\mathcal{U}$ and $\mathcal{I}$ denote the set of users and items, respectively. The fundamental task is to predict the next item $i_{t+1}$ or an item list that a user may prefer based on the user's chronological interaction history $H_u = (i_1, i_2, \dots, i_t)$.

The core intuition behind \textsc{RRCM} is that LLMs possess substantial inherent world knowledge and reasoning capabilities acquired during pre-training. For popular items or clear sequential patterns, the LLM can often infer the user's preference using only lightweight textual signals like item titles, and make accurate recommendations without additional evidence (e.g., a user watching the \emph{Harry Potter} series in order). In such cases, retrieving external context is unnecessary, computationally costly, and may even distract the model from the most relevant preference pattern.

However, effective recommendation may also require evidence that is not fully contained in the LLM's parametric memory. We identify two major sources of missing but valuable evidence:
\begin{enumerate}
    \item \textbf{Collaborative Memory Gap:} LLMs lack direct access to large-scale historical user--item interaction data, i.e., collaborative filtering signals, which are essential for capturing latent behavioral patterns beyond textual or semantic similarity.
    \item \textbf{Meta Memory Gap:} Including all item metadata into the context can result in prohibitive context length and inference costs. On the other hand, for niche, obscure, or newly introduced items absent from pretraining data, the LLM may lack the necessary attribute-level knowledge, leading to hallucinations if it's forced to reason without evidence.

\end{enumerate}

\begin{figure}[t]
\centering
\begin{minipage}{0.75\linewidth}
    \centering
    \includegraphics[width=0.8\linewidth]{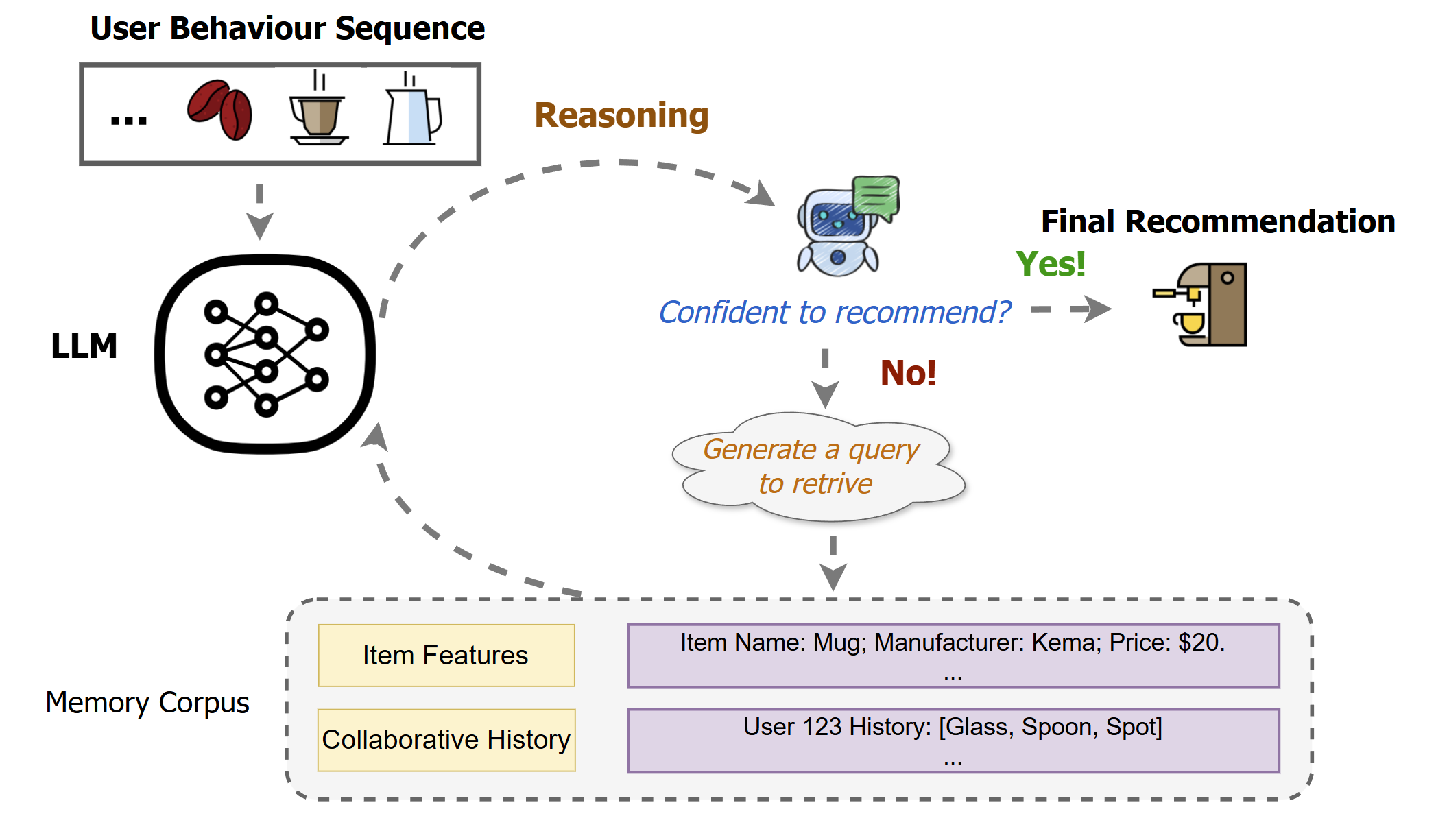}
\end{minipage}
\caption{Workflow of \textsc{RRCM} for ranking-driven retrieval over collaborative and meta memories.}
\vspace{-1mm}
\label{fig:rrcm}
\end{figure}

To address these challenges, we first define the model's default observation as a \emph{lightweight context} consisting primarily of item titles:
\begin{equation}
    S_{init} = [ \text{User Context}, \text{Title}(i_1), \dots, \text{Title}(i_t) ] .
\end{equation}
Then the objective is to learn a policy $\pi_\theta$ that determines whether the current context is sufficient for recommendation. When additional evidence is needed, the policy generates natural-language queries to retrieve from external recommender memories. In this way, retrieval is treated as an on-demand context construction action rather than a default preprocessing step with fixed rules.

\subsection{Dual-Memory Retrieval Corpus}\label{sec:corpurs}

We construct a unified retrieval corpus $\mathcal{M}$ that organizes heterogeneous recommendation evidence into a searchable textual format. Specifically, $\mathcal{M}$ contains two complementary recommender memories:
\begin{itemize}
    \item \textbf{Collaborative Memory:} Documents representing interaction sequences from historical users, e.g., \textit{``User 123 History: [Matrix, Inception, Interstellar, ...]''}. This provides behavioral evidence about how users with similar historical preferences continue their interactions.
    \item \textbf{Meta Memory:} Documents describing rich existing item metadata, e.g., \textit{``Movie Name: Inception; Director: Nolan; Genre: Sci-Fi; ...''}. This memory provides grounded item-level evidence, such as categories, creators, genres, brands, prices, and other features.
\end{itemize}

The retrieval engine $\mathcal{E}$ exposes a single flexible interface, \texttt{Retrieve($q$)}, where $q$ is a natural-language query generated by the policy. For instance, to retrieve item metadata, the model may ask \emph{``Find the director and genres of the movie `Primer'.''} To retrieve collaborative behavioral evidence, it may ask \emph{``Find histories of users who watched `Inception' and `Tenet'.''} This unified interface allows the LLM to access both behavioral and item-side evidence in natural language, while preserving its inherent language understanding and reasoning ability.

\subsection{Interleaved Reasoning and Memory Retrieval}\label{sec:retr}

We model inference as a multi-turn loop of reasoning and memory access. Given a current state $s_k$, which contains the initial user context and any previously retrieved evidence, the policy $\pi_\theta(a_k \mid s_k)$ generates a token trajectory $\tau$. The trajectory consists of a dynamic cycle:

\begin{enumerate}
    \item \textbf{Reasoning and Assessment (\texttt{<think>}):}
    The model analyzes the current state $s_k$ to infer the user's preference and assess whether the available evidence is sufficient for a reliable ranking decision. For example, it may determine whether the title-only history already reveals a clear pattern, or whether additional collaborative or item-side evidence is needed.

    \item \textbf{Action Decision:}
    \begin{itemize}
        \item \textbf{Memory Retrieval Action (\texttt{<tool\_call>}):} If the model requires additional evidence, it generates a query $q$ and calls \texttt{Retrieve($q$)} over $\mathcal{M}$. The retrieval engine returns the most relevant passage $d$, and the state is updated: $s_{k+1} \leftarrow s_k \oplus  d $. 
        The model then resumes reasoning over the updated context. This loop can repeat multiple times, enabling multi-step context construction, such as retrieving item metadata to disambiguate candidate items and then retrieving collaborative histories to identify behaviorally supported recommendations.

        \item \textbf{Termination Action (\texttt{<answer>}):} If the model judges that current context is sufficient, it stops retrieval/reasoning, and generates the final recommendation $\hat{y}$.
    \end{itemize}
\end{enumerate}

This iterative design allows \textsc{RRCM} to expand the context only when additional evidence is expected to improve the recommendation decision. The prompt template is provided in Appendix~\ref{sec:template}.

\subsection{Ranking-Driven Policy Optimization}

Since the optimal retrieval timing, memory type, and query formulation are unknown, we optimize the policy using reinforcement learning based directly on final recommendation quality. The key idea is to train the model not merely to retrieve relevant-looking evidence, but to retrieve evidence that directly improves the final ranking outcome.

\subsubsection{Generation Grounding}\label{sec:gene_ground}

The policy outputs a text string $\hat{y}$ for the predicted item title. To evaluate this output against the ground truth, we map it to the discrete item space $\mathcal{I}$. Following previous generative recommendation studies~\citep{gao2025sprec, bao2025bi}, we employ a semantic embedding approach. Let $E(\cdot)$ be a pre-trained sentence encoder (e.g., the Sentence Transformer~\citep{reimers-2019-sentence-bert}), we compute the similarity between $\hat{y}$ and all item titles in the candidate pool, the retrieve a recommendation list $\mathcal{L}_{cand}$ of the top-$N$ nearest items:
\begin{equation}
    \mathcal{L}_{cand} =
    \text{Top-N}_{i \in \mathcal{I}}
    \left(
    \frac{
    E(\hat{y}) \cdot E(\text{Title}_i)
    }{
    \|E(\hat{y})\| \|E(\text{Title}_i)\|
    }
    \right) .
\end{equation}
This grounding step converts the LLM's natural-language output into actionable recommendations from the item catalog, while retaining the flexibility of generative prediction.

\subsubsection{Ranking-Driven Reward Formulation}

We design a reward function $R(\tau)$ that reflects both ranking accuracy and structural validity. Specifically, recommendation accuracy is quantified as a weighted sum of top-$n$ ranking indicators on the grounded candidate list $\mathcal{L}_{cand}$ compared with the ground-truth item $i_{gt}$.

Let $\mathcal{N} = \{n_1, n_2, \dots, n_m\}$ be a set of rank cutoffs, e.g., $\{1, 5, 10, 50, 100\}$, and let $w_n$ be the corresponding weight for each cutoff. We use $\text{InTop}@n(\mathcal{L}_{cand}, i_{gt})$ to indicate whether $i_{gt}$ appears in the top-$n$ positions of $\mathcal{L}_{cand}$. The total reward is:
\begin{equation}
    R(\tau) =
    \left(
    \sum_{n \in \mathcal{N}}
    w_n \cdot \text{InTop}@n(\mathcal{L}_{cand}, i_{gt})
    \right)
    + \lambda \cdot \mathbb{I}_{parse},
\end{equation}
where $\mathbb{I}_{parse}$ is $0$ if the model follows the required \texttt{<answer>} format and $-1$ if parsing fails. This reward directly encourages the policy to construct contexts that improve the final top-$n$ ranking, while maintaining valid output structure.

\subsubsection{Optimization Instantiation with GRPO}

We instantiate the ranking-driven objective with token-level Group Relative Policy Optimization
(GRPO), which estimates the baseline from group rewards instead of using a separate critic. For each
input history $x$, we sample a group of $G$ outputs $\{y_1,\dots,y_G\}$, where each output may
include reasoning, retrieval actions, and the final recommendation.
The objective applies PPO-style clipping at the token level while constraining the policy to stay close
to a reference policy $\pi_{\mathrm{ref}}$:
\begin{equation}
\begin{split}
    \mathcal{J}(\theta)=
    \mathbb{E}_{x\sim\mathcal{D}}
    [
    \frac{1}{G}\sum_{i=1}^{G}\frac{1}{|y_i|}
    \sum_{t=1}^{|y_i|}
    (
    \min(
    \frac{\pi_\theta(y_{i,t}|x,y_{i,<t})}
    {\pi_{\mathrm{old}}(y_{i,t}|x,y_{i,<t})} A_i,
    \mathrm{clip}(
    \frac{\pi_\theta(y_{i,t}|x,y_{i,<t})}
    {\pi_{\mathrm{old}}(y_{i,t}|x,y_{i,<t})},
    1-\epsilon,
    \\ 1+\epsilon )
     A_i )
    -\beta \mathbb{D}_{\mathrm{KL}}
    (
    \pi_\theta(\cdot|x,y_{i,<t})
    \| \pi_{\mathrm{ref}}(\cdot|x,y_{i,<t})
    )
    )
    ].
\end{split}
\end{equation}
Here, $A_i$ is computed by standardizing the final rewards within the sampled group. We also mask the tokens of the retrieved documents in the output for training. Through this
objective, \textsc{RRCM} learns a ranking-driven memory-reading policy that retrieves collaborative
or item-side memory only when it improves final recommendation quality.

\section{Experiments}\label{sec:exp}

In this section, we introduce our experiments on three public e-commerce datasets, aiming to address the following research questions: \textbf{RQ1.} How does RRCM compare with prior methods for fine-tuning LLMs on recommendation tasks? \textbf{RQ2.} How do the individual components, including retrieving collaborative memory, item-metadata memory, and explicitly reasoning on them, contribute to RRCM’s performance? \textbf{RQ3.} How does RL fine-tuning shift the decision-making process for retrieval and reasoning strategies, such as retrieval frequency and context control?

\subsection{Experimental Setup}\label{sec:exp_set}

% Goodreads\footnote{\href{https://mengtingwan.github.io/data/goodreads}{https://mengtingwan.github.io/data/goodreads}}, MovieLens\footnote{\href{https://grouplens.org/datasets/movielens/}{https://grouplens.org/datasets/movielens/}},  and the \emph{CDs \& Vinyl} category from the Amazon Reviews\footnote{\href{https://jmcauley.ucsd.edu/data/amazon/index_2014.html/}{https://jmcauley.ucsd.edu/data/amazon/index\_2014.html}} dataset.

\textbf{Datasets.}
We evaluate on three public benchmarks: 
Goodreads \citep{wan2018item}, MovieLens \citep{harper2015movielens},  and the \emph{CDs \& Vinyl} category from Amazon Review dataset \citep{he2016ups}.
Following prior work \citep{bao2024decoding, gao2025sprec}, we remove users whose interaction sequences contain fewer than 10 events.
We then split the remaining data into training/validation/test sets with an 8:1:1 ratio.
To match the recent SPRec protocol for LLM-based recommendation \citep{gao2025sprec} for direct comparison, we further subsample 4,096 interactions from the training split for training, 512 from the validation split for validation, and 1,000 from the test split for testing. Then we construct the {collaborative-history retrieval corpus} by sampling from the remaining interactions after \textbf{excluding all interactions used in training/validation/test}.
This yields 50,000 records for MovieLens, 133,056 records for Goodreads, and 76,287 records for {CDs \& Vinyl}.

As described in Section \ref{sec:method}, we represent user histories using {item titles only} for both training, inference, and the collaborative-history memory to constrain the context length.
In addition, we build the {item-metadata memory} over all items in each dataset, which serves as the retrieval corpus for item-side metadata.
Specifically, Goodreads metadata includes \texttt{author}, \texttt{genres}, \texttt{series}, and \texttt{series\_id}; MovieLens metadata includes \texttt{director}, \texttt{all\_genres}, and \texttt{main\_genre}; and Amazon {CDs and Vinyl} metadata includes \texttt{price}, \texttt{salesRank}, \texttt{brand}, and \texttt{categories}.

\textbf{Evaluation Setting and Metrics.}
We evaluate top-$N$ recommendation quality using two standard information retrieval metrics.
Hit Ratio (HR@$N$) measures the fraction of test cases in which the ground-truth item appears within the top-$N$ ranked list \citep{hidasi2015session}. Normalized Discounted Cumulative Gain (NDCG@$N$) additionally accounts for the position of the ground-truth item, assigning higher weight to items ranked near the top \citep{kang2018self}.

As described in Section \ref{sec:gene_ground}, to leverage LLMs in generative recommendation, we
convert the LLM-generated output into a semantic embedding and match it against all items in the candidate pool to get the top-$N$ ones to form the final ranked candidate list.
This evaluation protocol follows recent LLM-based recommender work like BIGRec \citep{bao2025bi} and SPRec \citep{gao2025sprec}, ensuring fair and consistent comparisons. 
We use Qwen3-1.7B as the default LLM backbone, and set the parameters for reward formulation as $\mathcal{N} = (1, 5, 10, 50, 10, 100)$, ${w_n} = (0.5, 0.3, 0.1, 0.08, 0.02)$, and $\lambda=1$.
See Appendix \ref{sec:expdetail} for additional experimental details. 
% We also provide the prompt template and example dialogues for recommendation in Appendix \ref{sec:template}.

\textbf{Baselines.}
We compare  RRCM with both traditional sequential recommendation methods,
LLM-based recommenders trained with supervised fine-tuning, preference-alignment methods
based on DPO, reasoning-oriented LLM recommenders trained via SFT and/or RL, retrieval-augmented
LLM recommendation, and methods that explicitly incorporate collaborative filtering signals.

As a strong non-LLM baseline, we adopt \textbf{SASRec} \citep{kang2018self}, a self-attention based
sequential recommender. For standard SFT, we include \textbf{BIGRec} \citep{bao2025bi}, a principled
instruction-tuned LLM framework for sequential recommendation, and Debiasing-Diversifying Decoding (\textbf{D3}) \citep{bao2024decoding}, which
improves diversity and mitigates popularity bias.

For DPO-based alignment, we consider \textbf{DMPO} \citep{bai2024aligning}, which applies DPO by treating
sampled negatives as rejected responses, and the recent \textbf{SPRec} \citep{gao2025sprec}, which alternates SFT and DPO to improve the fairness and accuracy of user preference estimation.

For reasoning-oriented methods, we compare with \textbf{REPR} \citep{tsai2024leveraging} and \textbf{LatentR3} \citep{zhang2025reinforced}.
REPR investigates how explicit reasoning traces can benefit LLM recommendation under both
training-free prompting and SFT settings. LatentR3 introduces a latent reasoning stage before
producing the final recommendation, and optimizes the model with an SFT warm-up followed by
\textbf{GRPO-based RL} to better align reasoning with downstream recommendation accuracy.

We also include \textbf{K-RagRec} \citep{wang2025knowledge}, which augments LLM recommendation with
item-feature retrieval using a predefined (static) retrieval strategy on a knowledge graph, serving as a RAG-style baseline that does not learn when or what to retrieve. Finally, we compare against two CF-injection
approaches, \textbf{BinLLM} \citep{zhang2024text} and \textbf{CoLLM} \citep{zhang2025collm}. BinLLM encodes other users' interaction histories with binary string representation to fit the LLM input interface, whereas CoLLM maps collaborative embeddings into the LLM's latent space as additional conditioning. Note that we exclude some recent agentic recommendation systems like MemRec \citep{chen2026memrec} for performance comparison here, as they are training-free and rely on prompting/in-context demonstrations with much larger, often proprietary backbones (e.g., GPT-4o). In contrast, RRCM and our main baselines use smaller open-source backbones for controlled fine-tuning and practical training efficiency. See Appendix \ref{sec:exp_dis} for further discussion.

\begin{table}[t]
\centering
\caption{Performance comparison of RRCM (green) against prior SFT-based (yellow), DPO-based (gray), reasoning-based (brown) methods, and approaches incorporating metadata or collaborative signal retrieval (blue). HR@N and NG@N denote Hit Ratio@N and NDCG@N.}
\label{tab:main}
\small
\setlength{\tabcolsep}{4pt}
\renewcommand{\arraystretch}{1.08}
\resizebox{\linewidth}{!}{%
\begin{tabular}{lcccccccccccc}
\toprule
\multirow{2}{*}{Model}
& \multicolumn{4}{c}{Goodreads}
& \multicolumn{4}{c}{MovieLens}
& \multicolumn{4}{c}{Amazon CDs \& Vinyl} \\
\cmidrule(lr){2-5} \cmidrule(lr){6-9} \cmidrule(lr){10-13}
& HR@5 & NG@5 & HR@10 & NG@10
& HR@5 & NG@5 & HR@10 & NG@10
& HR@5 & NG@5 & HR@10 & NG@10 \\
\midrule

SASRec
& 0.0083 & 0.0057 & 0.0145 & 0.0076
& 0.0052 & 0.0042 & 0.0061 & 0.0045
& 0.0029 & 0.0028 & 0.0047 & 0.0034 \\

\midrule
\rowcolor{sftbg} 
BIGRec
& 0.0142 & 0.0113 & 0.0204 & 0.0159
& 0.0067 & 0.0063 & 0.0146 & 0.0104
& 0.0050 & 0.0043 & 0.0074 & 0.0068 \\

\rowcolor{sftbg} 
D3
& 0.0167 & 0.0130 & 0.0254 & 0.0192
& 0.0079 & 0.0075 & 0.0164 & 0.0121
& 0.0068 & 0.0047 & 0.0092 & 0.0085 \\

\midrule
\rowcolor{retrbg} 
DMPO
& 0.0159 & 0.0122 & 0.0237 & 0.0176
& 0.0090 & 0.0072 & 0.0166 & 0.0126
& 0.0070 & 0.0037 & 0.0079 & 0.0074 \\

\rowcolor{retrbg} 
SPRec

& 0.0194 & 0.0164 & 0.0256 & 0.0184
& 0.0094 & 0.0093 & \underline{0.0199} & 0.0130
& 0.0079 & 0.0068 & 0.0099 & 0.0075 \\

\midrule
\rowcolor{ppink} 
REPR
& 0.0178 & 0.0119 & 0.0220 & 0.0167
& 0.0084 & 0.0080 & 0.0156 & 0.0123
& 0.0051 & 0.0047 & 0.0088 & 0.0080 \\

\rowcolor{ppink} 
LatentR3
& 0.0188 & 0.0156 & 0.0281 & 0.0207
& \underline{0.0123} & 0.0103 & 0.0185 & 0.0141
& 0.0081 & 0.0062 & 0.0108 & 0.0095 \\

\midrule
\rowcolor{dpobg} 
K-RagRec
& \underline{0.0198} & 0.0168 & \underline{0.0309} & \underline{0.0233}
& 0.0118 & \underline{0.0113} & 0.0191 & \underline{0.0145}
& 0.0077 & 0.0064 & 0.0104 & 0.0089 \\

\rowcolor{dpobg}
CoLLM
& 0.0188 & 0.0156 & 0.0283 & 0.0230
& 0.0102 & 0.0088 & 0.0178 & 0.0132
& 0.0083 & 0.0073 & 0.0111 & 0.0090 \\

\rowcolor{dpobg}
BinLLM
& 0.0191 & \underline{0.0176} & 0.0298 & 0.0227
& 0.0110 & 0.0092 & 0.0194 & 0.0140
& \underline{0.0086} & \underline{0.0077} & \underline{0.0119} & \underline{0.0098} \\

\midrule
\rowcolor{rrcmbg}
RRCM
& \textbf{0.0215} & \textbf{0.0211} & \textbf{0.0364} & \textbf{0.0264}
& \textbf{0.0136} & \textbf{0.0130} & \textbf{0.0221} & \textbf{0.0160}
& \textbf{0.0096} & \textbf{0.0090} & \textbf{0.0144} & \textbf{0.0107} \\

\midrule
Improv.
& +8.59\% & +19.89\% & +17.80\% & +13.30\%
& +10.57\% & +15.04\% & +11.06\% & +10.34\%
& +11.63\% & +16.88\% & +21.01\% & +9.18\% \\
\bottomrule

\end{tabular}
}
\vspace{-2mm}
\end{table}

\subsection{Performance Comparison (RQ1)}\label{sec:performance}

We begin by comparing {RRCM} with all baselines on top-$N$ recommendation.
The results are summarized in Table~\ref{tab:main}.
Overall, we observe a consistent advantage of LLM-based recommenders over the traditional sequential baseline SASRec in our experimental setting, in line with the conclusions of previous research on LLMs for recommendation \citep{gao2025sprec, bao2025bi}.

% The significant performance gap suggests that leveraging LLMs, which have strong instruction-following ability, richer semantic priors, and general reasoning capability, is beneficial to improve the user preference understanding for recommendation task, especially when the training data size is relatively small. 

The direct SFT baselines, {BIGRec} and {D3}, consistently improve over SASRec but lag behind stronger LLM-based approaches. This indicates that simply fine-tuning an LLM to
imitate limited offline interaction patterns is insufficient to capture subtle preference cues and item-specific signals, especially when information  like collaborative filtering signals and rich item metadata is not present.

DPO-based methods yield additional improvements over pure SFT, particularly for {SPRec}, which alternates between SFT and DPO to iteratively debias recommendations.
 By constructing preferred vs.\ rejected pairs, DPO-style training encourages the model to
better distinguish good recommendations from poor ones, improving preference alignment and
ranking quality. However, these methods still do not explicitly incorporate collaborative information or
rich item metadata.
% , which limits their ability to resolve ambiguity and challenging cases. 
Moreover, the contrastive data construction for DPO can be problematic in recommendation: at each time step we only observe feedback for the single exposed target item, so the ``negative'' items sampled from the candidate
pool are often {unobserved} rather than truly disliked. Treating such possibly false negatives as rejected
responses can introduce biased supervision and may mislead the model’s preference learning.

Reasoning-oriented baselines such as {REPR} and {LatentR3} instruct the LLM to generate a
reasoning trace (explicit or latent) before outputting the final recommendation. While they
outperform SFT baselines, the improvements are generally modest. We argue this is expected:
reasoning without acquiring additional evidence mainly reorganizes the model’s internal priors to speculate for given downstream tasks. Unlike domains such as coding or math,
where abundant high-quality reasoning data exists in pretraining corpora, large-scale user behavioral data
from commercial recommender systems is typically proprietary and absent from public corpora. 
Therefore, the LLMs often lack enough domain knowledge to reason for recommendations.
As a result, directly prompting or fine-tuning LLMs for reasoning alone may provide limited gains.

Finally, {K-RagRec} (item-feature retrieval) and {BinLLM} (collaborative-signal injection) achieve the strongest performance among the compared baselines across datasets/metrics,
highlighting the value of exposing LLMs to item metadata and cross-user behavioral structure.
Nevertheless, both remain limited by {static} designs. K-RagRec relies on handcrafted retrieval rules (e.g., retrieving metadata only for the least popular items), which is crude and can overlook more nuanced scenarios. For instance, an item may be popular in the dataset yet still poorly understood by the LLM, such as platform-specific or niche content, and would therefore still benefit from metadata retrieval.
 Similarly, BinLLM (and related approaches such as CoLLM)
inject collaborative information via fixed representations (binary/text-like encodings or latent embedding
mapping), rather than allowing the LLM to request and interpret collaborative evidence in natural
language. These fixed schemes overlook a more flexible and intelligent strategy in which the model
\emph{learns} when collaborative/meta memory is needed and how to acquire the most
decision-relevant information for each instance.

Compared with all baselines, our proposed  {RRCM} consistently achieves the best performance across the
three datasets. RRCM learns a unified RL policy that
jointly decides \emph{whether} retrieval is needed, \emph{what} to retrieve, and \emph{how} to leverage the retrieved memory into the reasoning process to
produce improved recommendations. Crucially, these behaviors are learned end-to-end from a single
outcome objective for maximizing final ranking performance, without any manual rules for retrieval or
handcrafted mechanisms for injecting outside information. This enables RRCM to adaptively retrieve the most beneficial decision-relevant information and to generate more grounded, instance-specific
reasoning that better aligns recommendations with user preferences.

\begin{table}[t]
\centering
\caption{Performance comparison of RRCM (green) with three ablated variants: {RRCM w/o CF}, {RRCM w/o META}, and {RRCM w/o RE}, each removing a key component (gray).}
\label{tab:abl}
\small
\resizebox{1.02\linewidth}{!}{%
\begin{tabular}{l|cccc|cccc|cccc}
\toprule
\multirow{2}{*}{Model} 
& \multicolumn{4}{c|}{Goodreads} 
& \multicolumn{4}{c|}{MovieLens} 
& \multicolumn{4}{c}{Amazon CDs \& Vinyl} \\
\cmidrule(lr){2-5} \cmidrule(lr){6-9} \cmidrule(lr){10-13}
& HR@5 & NG@5 & HR@10 & NG@10 
& HR@5 & NG@5 & HR@10 & NG@10
& HR@5 & NG@5 & HR@10 & NG@10\\
\midrule

\rowcolor{retrbg} 
RRCM w/o CF    

& 0.0196 & \underline{0.0193} & 0.0331 & \underline{0.0241}
& 0.0119 & 0.0112 & 0.0199 & 0.0140
& 0.0081 & 0.0075 & 0.0112 & 0.0085 \\

\rowcolor{retrbg} 
RRCM w/o META  

& 0.0194 & 0.0187 & 0.0312 & 0.0227
& \underline{0.0124} & \underline{0.0118} & \underline{0.0201} & \underline{0.0145}
& 0.0086 & 0.0081 & \underline{0.0124} & \underline{0.0094} \\

\rowcolor{retrbg} 
RRCM w/o RE    
& \underline{0.0205} & 0.0188 & \underline{0.0333} & {0.0232}
& 0.0120 & 0.0116 & 0.0177 & 0.0136
& \underline{0.0090} & \underline{0.0084} & 0.0114 & 0.0092 \\
\midrule

\rowcolor{rrcmbg}
RRCM
& \textbf{0.0215} & \textbf{0.0211} & \textbf{0.0364} & \textbf{0.0264}
& \textbf{0.0136} & \textbf{0.0130} & \textbf{0.0221} & \textbf{0.0160}
& \textbf{0.0096} & \textbf{0.0090} & \textbf{0.0144} & \textbf{0.0107} \\
\bottomrule
\end{tabular}
}
\vspace{-3mm}
\end{table}

\subsection{Ablation Study (RQ2)}\label{sec:ablation}

In this section, we conduct ablation studies to examine the effect and contribution of the key components in RRCM. Specifically, we consider three variants: {RRCM w/o CF}, {RRCM w/o META}, and {RRCM w/o RE}. {RRCM w/o CF} is constructed by removing all interaction records of historical users from the retrieval memory corpus. We also modify the prompt to remove the instruction for the LLM to retrieve collaborative filtering information. All other components, including item metadata retrieval and reasoning mechanism, remain unchanged from the original RRCM. Similarly, {RRCM w/o META} is constructed by removing item metadata (i.e., the features) retrieval while keeping all other components unchanged. The last {RRCM w/o RE} keeps both collaborative filtering information and item metadata in the retrieval corpus. However, the LLM is not explicitly prompted to reason over the retrieved information. Instead, it is instructed to iteratively retrieve information until it is confident enough, and then directly make the recommendation.

The results in Table~\ref{tab:abl} show that {each} component of RRCM contributes consistently across all three datasets, validating our design of jointly integrating collaborative-history retrieval, item-metadata retrieval, and LLM reasoning within a unified RL optimization objective. We also observe dataset-specific patterns in component importance: removing metadata retrieval hurts most on {Goodreads}, removing reasoning hurts most on {MovieLens}, and removing collaborative-history retrieval hurts most on {CDs \& Vinyl}. This suggests that the bottleneck
differs by domain. Book titles in Goodreads can be ambiguous and many items are long-tail or old, making metadata crucial for disambiguation and grounding. In contrast, most of the CDs on Amazon or movies on MovieLens are relatively recent and well-covered by  LLM’s prior knowledge, while user taste can be very diverse and pattern-driven; thus collaborative histories and reasoning are more important for capturing preference regularities.

These findings further highlight the advantage of RRCM over methods with static retrieval rules or fixed CF-injection mechanisms. Rather than relying on handcrafted heuristics, RRCM learns an adaptive policy for retrieval, reasoning, and recommendation, which can automatically adjust to the characteristics of different datasets and application scenarios. We further compare RRCM with an ablation that always incorporates all evidence as the context. See discussions in Appendix \ref{app:all_evidence_comparison}.

% that decides {when} and {what} to retrieve and {how} to use the evidence in reasoning within a holistic RL framework, automatically adjusting to the characteristics of different datasets and application scenarios. We further compare RRCM with an ablation that always incorporates all evidence as the context. See relevant discussions in Appendix \ref{app:all_evidence_comparison}.

\subsection{Policy Behavior Shifting (RQ3)}
\begin{figure}[t]
\centering
\begin{minipage}{0.49\linewidth}
    \centering
    \includegraphics[width=0.99\linewidth]{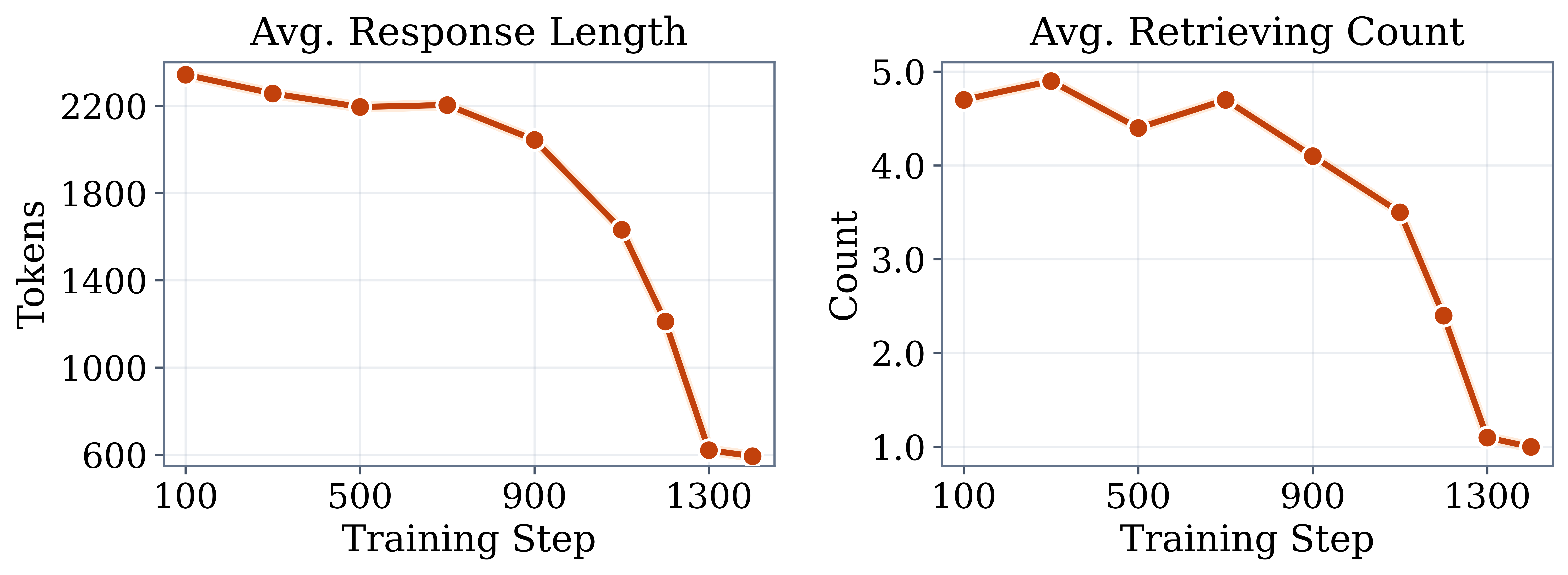}
\end{minipage}
\begin{minipage}{0.49\linewidth}
    \centering
    \includegraphics[width=0.99\linewidth]{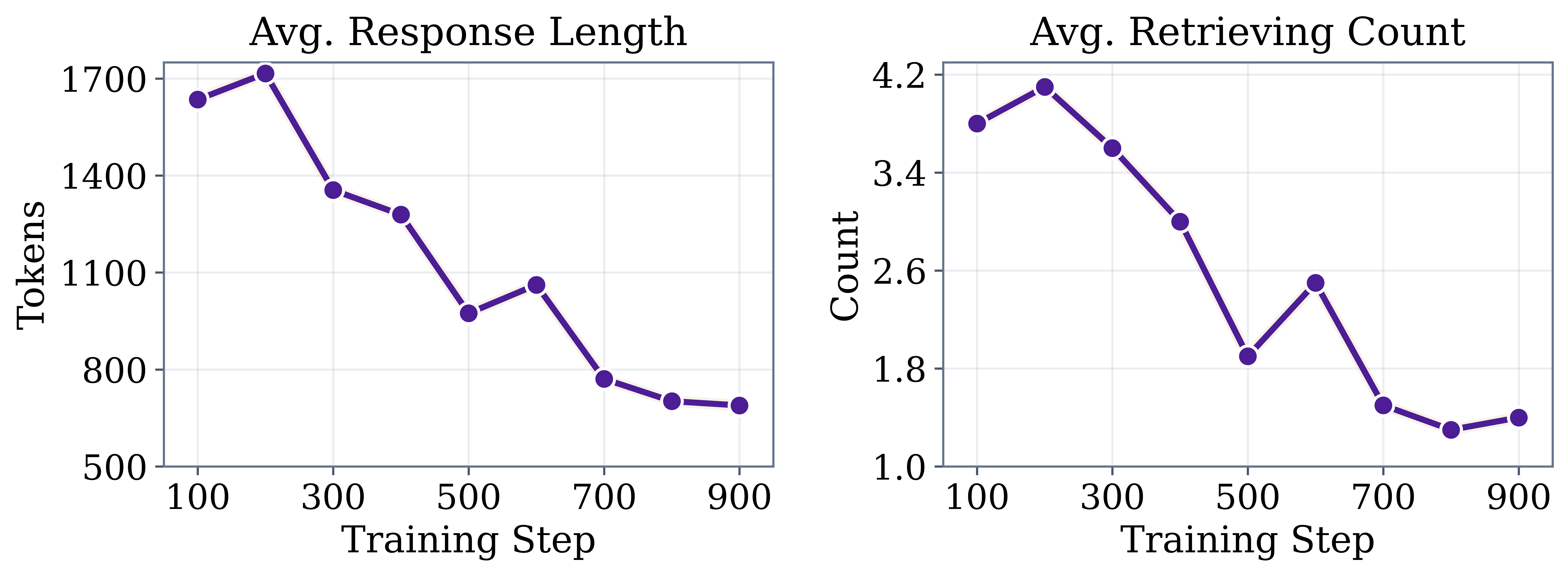}
\end{minipage}
\caption{ Average response token length and average retrieval count per recommendation during the training process until convergence on Goodreads (left) and CDs \&  Vinyl (right).}
\label{fig:bahavior}
\vspace{-2mm}
\end{figure}

To analyze how the RL policy evolves its retrieval strategy and reasoning behavior, we track two efficiency indicators throughout training: (i) the average number of retrieval calls per recommendation and (ii) the average response length (in tokens), including all generated queries, reasoning, and recommendation across the entire dialogue. Figure~\ref{fig:bahavior} reports these quantities, where each point aggregates statistics over training samples within a range of optimization steps (e.g., 300--500 steps).

As training progresses, both the retrieval operation counts and the response length exhibit a steady decline. This suggests a clear policy behavior shift: early in training, the agent explores by retrieving more evidence and producing longer reasoning traces; later, it learns to make recommendations with fewer retrieval actions and more concise responses. That is, the policy becomes increasingly selective, retrieving collaborative histories or item metadata only when additional evidence is necessary, and otherwise relying on the existing context and the LLM’s internal knowledge for recommendation.

These results indicate that RRCM learns an {on-demand} information acquisition strategy that balances accuracy and context cost. By avoiding static, rule-based retrieval heuristics and instead learning when retrieval is beneficial, RRCM mitigates the context-length bottleneck while still leveraging external evidence when it meaningfully improves recommendation quality. We present computation cost analysis in Appendix \ref{app:computation_cost}, and more model behavior analysis in Appendix \ref{app:model_behavior} and \ref{app:scaling_behavior}.

\section{Conclusion}

This paper proposes \textsc{RRCM}, an adaptive retrieval-and-reasoning framework for LLM-based recommendation that addresses two limitations of existing approaches: collaborative signals are only implicitly captured, and rich item metadata is costly to include for long histories. Starting from a lightweight user-history context, \textsc{RRCM} learns whether to retrieve, which memory (collaborative or meta) to query, and how to use the evidence, optimized via a GRPO-based ranking objective that directly rewards top-$k$ recommendation quality. Experiments on three public datasets show \textsc{RRCM} outperforms strong sequential, LLM-based, retrieval-augmented, and CF-enhanced baselines, with ablations confirming the complementary roles of both memories and reasoning.

% This paper studies ranking-driven context construction for LLM-based recommendation. We identify two limitations of existing LLM recommenders: collaborative signals are only implicitly captured from limited training contexts, while rich item metadata is costly to include for long histories. We propose \textsc{RRCM}, an adaptive retrieval-and-reasoning framework over collaborative and meta memories. Starting from a lightweight user-history context, \textsc{RRCM} learns whether to retrieve, which memory to query, and how to use the acquired evidence, with its policy optimized by a GRPO-based ranking objective that directly rewards final top-$k$ recommendation quality.

% Experiments on three public datasets show that \textsc{RRCM} outperforms strong sequential, LLM-based, retrieval-augmented, and CF-enhanced baselines. Ablations and policy analyses further verify the complementary roles of both memories and reasoning, and show that \textsc{RRCM} learns selective retrieval for accurate and context-efficient recommendation.

\subsection*{AI use statement}

In this work, we used generative AI tools for implementing methods.
We have not used generative AI tools to help develop theoretical models or conceptual frameworks, formulate mathematical claims, provide critical ingredients for proving mathematical claims, propose or refine hypotheses, design or provide feedback on research  methodology or experiments,  assist with translation, clean and reformat dataset, support qualitative and thematic data analysis, interpret results,
and other tasks to generate synthetic data sets, assist in the writing of proofs, are not applicable to this work.
Additionally, we used generative AI tools for editing a research paper to improve readability. We have reviewed all AI-assisted work. For example, LLM-generated code was carefully verified and tested for correctness. We take responsibility for the final content of this work, including text, claims or artifacts produced with the aid of generative AI.

\subsection*{Reproducibility statement}

We elaborate on the experiment details in the Section \ref{sec:exp} and Appendix \ref{sec:expdetail}, including datasets, baselines, hyper-parameters, etc. We also release the code and data for reproducibility. Our code and data are accessible at \url{https://anonymous.4open.science/r/RRCM-6233}.

\bibliography{reference}
\bibliographystyle{iclr2027_conference}

\newpage
\appendix

\newpage
\appendix

\section{Experiment Details}\label{sec:expdetail}

We conduct all experiments on a dedicated NVIDIA GH200 superchip equipped with an H100 GPU. Specifically, the NVIDIA H100 GPU provides 96 GiB of HBM3 memory, and all experiments are run on a system with 256 GB of RAM and a 1 TB SSD. For all models, we use the Adam optimizer and tune the learning rate within the range ([1e-7, 5e-7, 1e-6, 5e-6, 1e-5]). By default, we fine-tune
{Qwen3-1.7B} using GRPO \citep{grpo}. GRPO is configured with a KL coefficient of 0.001 and a clip ratio of 0.2.
% , and an SFT warm-up ratio of 0.285. 
During rollout, we use a group size of 8 and a sampling temperature of 1.0. We enable tool-calling via {SGLang} \citep{zheng2024sglang} to allow the LLM to generate retrieval queries and set maximal turns to 5. For retrieval, we use the {Flat E5} retriever \citep{wang2022text} following prior work \citep{jin2025search}. Flat indexing encodes the corpus into dense embeddings and retrieves relevant documents via efficient embedding matching in the embedding space. As illustrated in Section \ref{sec:retr}, the retriever returns the most relevant passage each time to constrain the context and maintain efficiency. If retrieval fails, the retriever returns a blank placeholder.

To compute the item embeddings for item matching and finalizing the ranking list as described in Section \ref{sec:gene_ground}, we use the Sentence Transformer \citep{reimers-2019-sentence-bert}. Note that this embedding-based mapping strategy is widely adopted in existing generative recommendation methods \citep{gao2025sprec, bao2025bi}, as discussed in Section~\ref{sec:gene_ground}. Following these works, we use nearest-neighbor similarity to map the LLM-generated output to items in the candidate pool when exact matching is unavailable. Since the focus of this paper is to study RL-based memory retrieval and reasoning for recommendation, we adopt the standard item-matching protocol from prior work, while referring readers to these studies for detailed discussions on some general issues of generative recommendation by LLMs, such as format failures and hallucinated item names.

For all baselines, we follow the original papers for implementation details and hyperparameter tuning. We run each experiment with three random seeds and report the mean performance. Statistical significance  is evaluated using a paired t-test with $p < 0.05$.

\section{Additional Experiments}\label{sec:add_exp}

\subsection{Experiments with Other Backbone LLM}\label{app:backbone}

In addition to using Qwen3-1.7B as the base model, we also evaluate with a larger backbone, Llama-3.2-3B. The corresponding results, reported in Table~\ref{tab:llama}, consistently demonstrate the robustness of RRCM and its sustained performance gains over the baseline methods. The results further highlight the cross-model transferability of the optimization benefits provided by the proposed RL-based retrieval-and-reasoning framework for recommendation.

Furthermore, compared with the results in Table~\ref{tab:main}, the performance advantage of
K-RagRec over BinLLM becomes less pronounced here. This is reasonable because K-RagRec mainly benefits from retrieving item metadata, which helps the LLM better understand item attributes. Such retrieval can be particularly useful for relatively small LLMs such as Qwen3-1.7B, which may encode less
world knowledge about niche or long-tail items. However, when using a larger backbone such as Llama-3.2-3B, the model may already possess broader knowledge of many items. As a result,
metadata retrieval becomes beneficial in fewer cases, reducing its relative contribution and narrowing the gap between K-RagRec and collaborative-signal retrieval methods such as BinLLM.

\begin{table}[h]
\centering
\caption{Performance comparison with Llama-3.2-3B on Goodreads. The best and second-best results are highlighted in bold and underlined, respectively.}
\label{tab:llama}
\resizebox{\linewidth}{!}{%
\begin{tabular}{c|c|cc|cc|ccc|c|c}
\toprule
 & SASRec & BIGRec & SPRec & REPR & LatentR3 & K-RagRec & CoLLM & BinLLM & RRCM & Improv. \\
\midrule
HR@5  & 0.0083 & 0.0155 & 0.0179 & 0.0155 & 0.0204 & \underline{0.0228} & 0.0217 & 0.0223 & \textbf{0.0248} & +8.77\% \\

NG@5  & 0.0057 & 0.0137 & 0.0168 & 0.0124 & 0.0166 & 0.0205 & 0.0197 & \underline{0.0211} & \textbf{0.0228} & +8.06\% \\

\midrule
HR@10 & 0.0145 & 0.0246 & 0.0306 & 0.0212 & 0.0283 & 0.0301 & 0.0295 & \underline{0.0312} & \textbf{0.0360} & +15.38\% \\

NG@10 & 0.0076 & 0.0209 & 0.0213 & 0.0167 & 0.0235 & \underline{0.0248} & 0.0239 & 0.0244 & \textbf{0.0267} & +7.66\% \\
\bottomrule
\end{tabular}%
}
\end{table}

\begin{figure}[t]
\centering
\begin{minipage}{0.85\linewidth}
    \centering
    \includegraphics[width=0.85\linewidth]{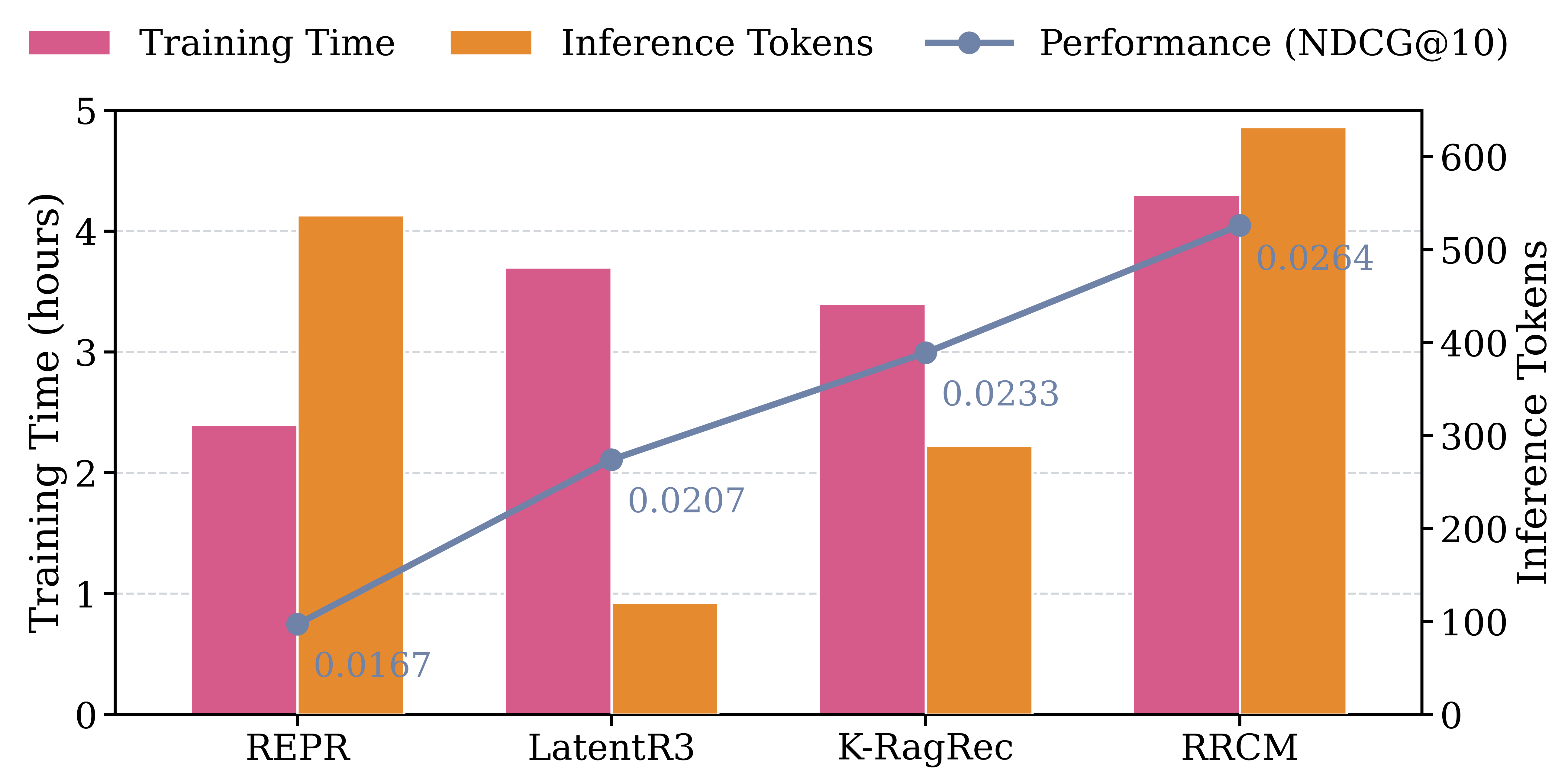}
\end{minipage}
\caption{Computation cost and performance comparison.}
\vspace{-1mm}
\label{fig:cost}
\end{figure}

\subsection{Computation Cost and Efficiency}
\label{app:computation_cost}

We evaluate RRCM's computational cost through training time,
inference token consumption, wall-clock latency, and memory-index
size. Our comparisons focus on our previous baselines that incorporate reasoning or retrieval: REPR, an SFT-based explicit-reasoning method;
LatentR3, an RL-based latent-reasoning method; and K-RagRec,
a static metadata-retrieval method.

\paragraph{Training cost and inference tokens.}
Figure~\ref{fig:cost} compares computational cost and recommendation performance
on Goodreads. Training costs reflect both the optimization procedure
and the operations performed during rollout generation. RL-based
methods incur additional overhead from sampling trajectories and
updating the policy. At inference time, we measure token consumption
including input context and generated outputs. LatentR3 consumes
the fewest tokens through its compact latent reasoning, while REPR
generates explicit reasoning and K-RagRec augments the context with
retrieved metadata.

RRCM incurs the highest training time and inference token consumption
among these methods because it combines explicit reasoning with
memory retrieval. This additional computation accompanies improved
recommendation quality: its NDCG@10 exceeds that of REPR and
LatentR3 by 58.1\% and 27.5\%, respectively. Similar trade-offs
arise in other agentic recommenders. For example, Table~5 of
MemRec \citep{chen2026memrec} paper reports approximately 6.8K total tokens
per user session for its standard configuration, compared with
1.2K for direct recommendation. These results reflect the common trade-off in LLM-based systems between computational cost and task performance: explicit reasoning and retrieval-augmented context can improve model capability, but they also introduce additional token consumption and retrieval latency \citep{wei2022chain, tsai2024leveraging, rag, jin2025search}. As a result, we argue that the additional cost is a general characteristic of agentic recommendation methods, rather than an issue unique to RRCM.

\paragraph{Wall-clock latency and memory footprint.}
Besides token counts and training time,
we further measure inference latency, retrieval
latency, and memory-index size for REPR, K-RagRec, and RRCM on
Goodreads and MovieLens, as reported in
Table~\ref{tab:inference_latency}.

The results are consistent: incorporating dual memories and multi-step reasoning introduces additional latency. However, this reflects the common performance-efficiency trade-off in retrieval-augmented and agentic LLM systems. Explicit evidence acquisition improves recommendation quality but requires additional generation and retrieval operations. Importantly, retrieval itself accounts for only a very limited portion of the total wall-clock latency; most inference time comes from autoregressive LLM generation.

% For reproducibility, specify whether retrieval latency is averaged
% per retrieval call or accumulated per recommendation instance.
% Also document the timing setup, including batch size and whether
% query encoding and tool-call overhead are included.

\begin{table}[t]
    \centering
    \caption{Inference latency and memory-index size on Goodreads
    and MovieLens. Dashes indicate that no external retrieval
    index is used.}
    \label{tab:inference_latency}
    \small
    \setlength{\tabcolsep}{9pt}
    \begin{tabular}{llrrr}
        \toprule
        & & \multicolumn{2}{c}{Latency} & Memory index \\
        \cmidrule(lr){3-4}
        Dataset & Model & Inference (s) & Retrieval (ms) & Size (GB) \\
        \midrule
        Goodreads & REPR     & 3.21 & -- & --   \\
                  & K-RagRec & 2.84 & 15 & 0.31 \\
                  & RRCM     & 4.08 & 19 & 0.52 \\
        \midrule
        MovieLens & REPR     & 3.02 & -- & --   \\
                  & K-RagRec & 2.35 & 12 & 0.13 \\
                  & RRCM     & 3.63 & 15 & 0.21 \\
        \bottomrule
    \end{tabular}
\end{table}

\paragraph{Selective retrieval and explicit cost control.}
RRCM allows the policy to recommend directly when the current
context is sufficient. Figure \ref{fig:bahavior} shows that retrieval frequency and generated response length decrease during training, indicating
that the policy learns to use fewer retrieval actions and shorter
responses under the original ranking-driven reward. This behavior
emerges without an explicit retrieval-cost penalty.

To examine whether inference cost can be reduced further, we evaluate a reward variant that adds an output-length penalty on Goodreads, defined as $-0.3 \times max(0, (OutputTokenLen - 256)/256)$, where a longer output than the 256-token reference incurs a larger penalty. Table~\ref{tab:cost_aware_reward} shows that the penalty
reduces average inference token consumption from 593 to 507
(14.5\%) and retrievals per instance from 0.93 to 0.82 (11.8\%).
Separately, average reasoning length decreases by approximately
16\%. These reductions accompany lower performance across all
four ranking metrics, demonstrating a trade-off between
recommendation quality and computational cost. The original reward
achieves higher ranking accuracy in this comparison, while the
penalized variant offers lower token consumption and retrieval
frequency.

% Add the exact output-length penalty definition, its coefficient,
% and any length normalization used in this experiment.

\begin{table}[t]
    \centering
    \caption{Effect of an output-length penalty on Goodreads.
    Inference tokens and retrieval counts are averaged per instance.
    NG denotes NDCG.}
    \label{tab:cost_aware_reward}
    \small
    \setlength{\tabcolsep}{5pt}
    \begin{tabular}{lrrrrrr}
        \toprule
        & \multicolumn{4}{c}{Recommendation quality}
        & \multicolumn{2}{c}{Inference cost} \\
        \cmidrule(lr){2-5}
        \cmidrule(lr){6-7}
        Reward setting & HR@5 & NG@5 & HR@10 & NG@10
                       & Tokens & Retrievals \\
        \midrule
        Original reward
            & 0.0215 & 0.0211 & 0.0364 & 0.0264 & 593 & 0.93 \\
        With cost penalty
            & 0.0209 & 0.0206 & 0.0339 & 0.0252 & 507 & 0.82 \\
        \bottomrule
    \end{tabular}
\end{table}

\paragraph{Practical deployment.}
Based on the computational cost and latency analysis above, we
argue that RRCM is most applicable to settings where improvements
in recommendation quality justify additional inference cost.
It is unlikely to replace the earliest retrieval stage
of a production recommender serving millions of requests, where
strict latency control and computational efficiency are critical.
A more practical option is to use RRCM as a later-stage reranker
over candidates produced by an efficient retrieval model.
In this setting, its adaptive policy can avoid unnecessary evidence
acquisition, while deployment costs can be further managed through
memory caching, batched approximate nearest-neighbor search, and
limits on retrieval turns or reasoning length. Selective activation
for uncertain or high-value requests provides another option for
balancing recommendation quality with serving efficiency.

\section{Expanded Related Work}\label{sec:exp_dis}

One core formulation of RRCM is to cast recommendation as an agentic, multi-turn process that
interleaves reasoning and retrieval, and to optimize this process with reinforcement learning. Here, we provide a more detailed discussion of recent work on agentic recommendation and RL-based alignment for LLM recommendation.

Recmind \citep{wang2024recmind}, Star \citep{lee2024star}, MemRec \citep{chen2026memrec}, and QUEREC \citep{han2025rethinking} represent recent efforts to systematically study agentic recommendation. However, their frameworks are entirely training-free, with agentic behavior primarily guided by in-context examples and general external knowledge with much larger, often proprietary backbones (e.g., GPT-4o and GPT-OSS-120B in MemRec). In contrast, RRCM and our main baselines, such as K-RagRec and CoLLM, use smaller open-source backbones (Qwen3-1.7B and Llama-3.2-3B) to enable controlled fine-tuning and practical training efficiency.

This difference makes direct and fair comparison difficult. 
When a training-free method such as MemRec is transferred to
its agentic behavior and the same 1.7B backbone,
\begin{wraptable}{rh}{0.55\textwidth}
\renewcommand\arraystretch{1.0}
\footnotesize
\centering
\caption{Comparison of MemRec with Qwen3-1.7B. }
\vspace{-1mm}
\label{tab:MemRec}
\begin{tabular}{ccccc}
\toprule
Model &HR@5	&NG@5	&HR@10	&NG@10 \\
\midrule
SASRec	&0.0083	&0.0057	&0.0145	&0.0076\\
MemRec	&0.0063	&0.0049	&0.0140	&0.0066\\
RRCM	&0.0215	&0.0211	&0.0364	&0.0264\\
\bottomrule
\end{tabular}
\vspace{-1mm}
\end{wraptable}
recommendation quality degrades substantially. To verify that, we implemented MemRec with the Qwen3-1.7B under the same data splits and evaluation protocol on Goodreads. As show in Table \ref{tab:MemRec}, it performs substantially worse than RRCM and also even underperforms SASRec in many cases.

 \cite{zhu2025collaborative} further explore agentic retrieval of collaborative information for LLM recommendation. Nevertheless, similar
to the works \citep{zhu2025llm,zhu2025collaborative} discussed in Section~\ref{sec:intro}, their setting focuses on conversational recommendation, where the system interacts with users over multiple natural-language turns. This is
different from the general sequential recommendation setting considered in our work.

RL-based alignment has also been explored for LLM recommendation. Rec-r1 \citep{lin2025rec} use RL to fine-tune LLMs for recommendation-related optimization, but the final recommendation is still produced by an existing classic recommender system; the LLM only provides auxiliary textual features such as rewritten queries, user profiles, or item descriptions. This differs from RRCM, where the LLM is
directly optimized to generate the final recommendation. Similarly, DeepRec \citep{zheng2025deeprec} employs a classic recommendation model for retrieval and requires process rewards, which is distinct from our end-to-end framework design.
Another recent work Rank-GRPO \citep{zhu2025rank}
applies RL to directly fine-tune LLMs for recommendation. However, it is also specifically designed for conversational recommendation, making it not directly comparable to RRCM.

% Compared with conventional LLM-based baselines, our method introduces additional inference cost from retrieving collaborative histories and item metadata, as well as from generating reasoning tokens. However, this reflects a common trade-off in LLM-based methods between recommendation quality and computational overhead \citep{rag, zhang2024text, tsai2024leveraging}. For example, chain-of-thought prompting generally improves task performance by producing intermediate reasoning that can significantly increase the inference time, while RAG-style methods provide valuable external evidence at the cost of longer contexts and retrieval latency. Similarly, RRCM uses retrieval and reasoning as additional computation to obtain more accurate and better-grounded recommendations, while learning to invoke them selectively when they are most beneficial.

% That said, we further compare RRCM with representative strong baselines in terms of both inference cost and recommendation performance. Specifically, following our main results in Table \ref{tab:main}, we select \textbf{SPRec} as the strongest non-reasoning LLM baseline, \textbf{LatentR3} as the latent-reasoning method, \textbf{REPR} as a direct reasoning method, \textbf{K-RagRec} as a metadata-retrieval method, and \textbf{BinLLM} as a collaborative-information injection method.

\subsection{Model Behavior Analysis}
\label{app:model_behavior}

Beyond the aggregate retrieval frequency and response length reported
in Section~4.4, we examine which memory sources RRCM selects, whether
retrieved evidence contributes to recommendation quality, how well
queries and retrieved documents align with their intended purpose,
and how frequently failures occur.

\paragraph{Memory-source selection.}
Table~\ref{tab:memory_source_selection} reports the proportions of retrieval actions directed toward collaborative and metadata memories. The learned policy exhibits distinct preferences across datasets:
\begin{wraptable}{r}{0.45\columnwidth}  % 双栏用 \columnwidth，单栏页面可改用 0.45\textwidth
\footnotesize
\centering
\setlength{\tabcolsep}{3pt}             % 缩小列间距（默认通常是 6pt）
\vspace{-2mm}
\caption{Memory-source selection ratios.}
\label{tab:memory_source_selection}
\begin{tabular}{lcc}
\toprule
Dataset & \shortstack{Collaborative\\memory} & \shortstack{Metadata\\memory} \\
\midrule
Goodreads    & 38.4\% & 61.6\% \\
CDs \& Vinyl & 73.3\% & 26.7\% \\
\bottomrule
\end{tabular}
\vspace{-3mm}
\end{wraptable}
metadata accounts for 61.6\% of retrievals on Goodreads, whereas
collaborative memory accounts for 73.3\% on CDs \& Vinyl.
This pattern agrees with the component ablations in Section~4.3,
where removing metadata retrieval causes the largest degradation on
Goodreads and removing collaborative retrieval is most harmful on
CDs \& Vinyl. These observations suggest that the policy's source
selection automatically learns and reflects the relative value of different evidence types
across domains.

\paragraph{Usefulness of retrieved evidence.}
To examine the contribution of retrieved information, we conduct an
evidence-removal diagnostic on test trajectories containing retrieval
actions. Specifically, we remove the retrieved context and instruct the model to
recommend directly while retaining reasoning. We then compare the
recommendation outcomes with those of the original trajectories.
Table~\ref{tab:evidence_usefulness} reports the improvement ratios
from this comparison across four ranking metrics.

\begin{table}[h]
    \centering
    \caption{Reported improvement ratios in the evidence-removal
    diagnostic.}
    \label{tab:evidence_usefulness}
    \small
    \setlength{\tabcolsep}{12pt}
    \begin{tabular}{lrrrr}
        \toprule
        Dataset & HR@5 & NG@5 & HR@10 & NG@10 \\
        \midrule
        Goodreads & 73.5\% & 90.5\% & 81.1\% & 97.7\% \\
        MovieLens & 80.4\% & 94.7\% & 84.9\% & 93.6\% \\
        \bottomrule
    \end{tabular}
\end{table}

\paragraph{Query quality and retrieval relevance.}
We use Claude Sonnet 4.6 as an LLM judge to evaluate query
appropriateness and retrieved document relevance separately.
For metadata queries, the judge assesses whether the requested
attributes are relevant to a historical item or an item under
consideration, given the current user interaction sequence.
For collaborative queries, it assesses whether the query appropriately
seeks behavioral evidence from users with similar preference
histories. The judge then evaluates whether the returned document
is relevant to the generated query.

As shown in Table~\ref{tab:query_retrieval_accuracy}, query accuracy
ranges from 82.4\% to 93.4\%, while retrieval accuracy ranges from
85.2\% to 94.5\%. Metadata queries and retrievals receive higher
accuracy scores than their collaborative counterparts on both
datasets. This difference suggests that identifying relevant
behavioral histories is more challenging than requesting and
retrieving item attributes. These automated assessments provide
additional diagnostics of retrieval behavior, alongside the
recommendation-based evaluations.

\begin{table}[t]
    \centering
    \caption{LLM-judged query accuracy and retrieval accuracy,
    separated by memory source.}
    \label{tab:query_retrieval_accuracy}
    \small
    \setlength{\tabcolsep}{9pt}
    \begin{tabular}{lcccc}
        \toprule
        & \multicolumn{2}{c}{Query accuracy}
        & \multicolumn{2}{c}{Retrieval accuracy} \\
        \cmidrule(lr){2-3}
        \cmidrule(lr){4-5}
        Dataset & Metadata & Collaborative
                & Metadata & Collaborative \\
        \midrule
        Goodreads & 93.4\% & 87.7\% & 94.5\% & 91.9\% \\
        MovieLens & 90.9\% & 82.4\% & 88.4\% & 85.2\% \\
        \bottomrule
    \end{tabular}
\end{table}

\paragraph{Sensitivity to retrieval noise.}
We further examine the effect of irrelevant evidence by replacing
the retrieved document with an unrelated document from the same
corpus with probability 0.5 for test cases containing retrieval.
Table~\ref{tab:retrieval_noise} compares the original and corrupted
settings. Corruption reduces performance across all four metrics
on both Goodreads and MovieLens, supporting the importance of
evidence relevance to RRCM's recommendation quality. The results
also identify retrieval quality as a practical limitation: errors
in evidence acquisition can propagate to the final recommendation.

\begin{table}[h]
    \centering
    \caption{Recommendation performance with original retrieval
    and with 50\% retrieval corruption.}
    \label{tab:retrieval_noise}
    \small
    \setlength{\tabcolsep}{9pt}
    \begin{tabular}{llrrrr}
        \toprule
        Dataset & Retrieval setting
                & HR@5 & NG@5 & HR@10 & NG@10 \\
        \midrule
        Goodreads & Original
                  & 0.0215 & 0.0211 & 0.0364 & 0.0264 \\
                  & 50\% corrupted
                  & 0.0173 & 0.0168 & 0.0334 & 0.0224 \\
        \midrule
        MovieLens & Original
                  & 0.0136 & 0.0130 & 0.0221 & 0.0160 \\
                  & 50\% corrupted
                  & 0.0112 & 0.0108 & 0.0183 & 0.0133 \\
        \bottomrule
    \end{tabular}
\end{table}

\paragraph{Sensitivity to catalog grounding.}
As described in Section \ref{sec:exp_set}, we follow BIGRec and SPRec in using Sentence-Transformer-based catalog grounding to map generated titles
to valid catalog items. While this approach preserves generative
flexibility, semantically similar or ambiguous titles may affect
the resulting rankings. We therefore additionally evaluate RRCM
and strong baselines using exact title matching on Goodreads by asking the model to generate 10 item titles every time.
As shown in Table~\ref{tab:exact_title_matching}, RRCM consistently
outperforms all compared methods across the four ranking metrics.
These results support the robustness of its gains under a stricter
catalog-matching criterion and suggest that the improvements are
not solely attributable to semantic-nearest-neighbor grounding.

\begin{table}[h]
    \centering
    \caption{Performance with exact title matching on Goodreads.}
    \label{tab:exact_title_matching}
    \small
    \setlength{\tabcolsep}{12pt}
    \begin{tabular}{lrrrr}
        \toprule
        Model & HR@5 & NG@5 & HR@10 & NG@10 \\
        \midrule
        BIGRec   & 0.0116 & 0.0091 & 0.0173 & 0.0134 \\
        SPRec    & 0.0147 & 0.0131 & 0.0235 & 0.0162 \\
        K-RagRec & 0.0168 & 0.0134 & 0.0240 & 0.0196 \\
        CoLLM    & 0.0159 & 0.0123 & 0.0218 & 0.0181 \\
        \midrule
        RRCM & \textbf{0.0189} & \textbf{0.0187} & \textbf{0.0304} & \textbf{0.0227} \\
        \bottomrule
    \end{tabular}
\end{table}

\paragraph{Output and retrieval reliability.}
Table~\ref{tab:behavior_failure_rates} reports hallucinated-title,
parse-failure, and retrieval-failure rates across all three datasets.
For this analysis, retrieval failure is assessed by Claude Sonnet 4.6
based on whether the returned document is relevant to the generated
query and its intended retrieval purpose; this criterion includes
semantically unsuccessful retrievals.

Parse failures are infrequent, ranging from 0.1\% to 0.4\%.
Hallucinated-title rates range from 2.5\% to 4.1\%, while
retrieval-failure rates range from 5.2\% to 8.8\%.
These results indicate that output formatting is generally reliable,
while title generation and evidence acquisition remain sources of
error. In particular, the retrieval failures complement the
corruption experiment by highlighting the value of improving
retrieval relevance.

\begin{table}[t]
    \centering
    \caption{Output and retrieval failure rates.
    Retrieval failures are judged by query relevance and intended
    retrieval purpose.}
    \label{tab:behavior_failure_rates}
    \small
    \setlength{\tabcolsep}{9pt}
    \begin{tabular}{lccc}
        \toprule
        Dataset & Hallucinated title & Parse failure & Retrieval failure \\
        \midrule
        Goodreads    & 3.7\% & 0.4\% & 6.0\% \\
        MovieLens    & 2.5\% & 0.1\% & 5.2\% \\
        CDs \& Vinyl & 4.1\% & 0.3\% & 8.8\% \\
        \bottomrule
    \end{tabular}
\end{table}

\subsection{Scaling Behavior Analysis}
\label{app:scaling_behavior}

We examine RRCM's behavior as the training set, evaluation set,
and retrieval corpus increase in size. These experiments assess
whether its performance advantage persists with additional training
data and broader evaluation, and how expanding the available
evidence affects recommendation quality and retrieval latency.

\paragraph{Scaling the training set.}
Our main experiments follow the training-data protocol adopted in
prior LLM-based recommendation studies, including SPRec \citep{gao2025sprec} and BIGRec \citep{bao2025bi}.
Using the same training set across methods enables controlled and
computationally feasible comparisons among fine-tuned LLM baselines.
However, traditional recommenders may benefit substantially from
additional interaction data. We therefore expand the training set
to three times its original size on Goodreads and train both SASRec and RRCM
under this expanded setting.

Table~\ref{tab:training_scaling} shows that RRCM retains a clear
advantage over SASRec across all four metrics. This result suggests
that its advantage over the traditional sequential baseline
persists beyond the original limited-data setting.

% Specify the dataset used for this experiment in the caption.
\begin{table}[h]
    \centering
    \caption{Performance with a training set three times the original
    size on Goodreads. Both methods are trained under the expanded setting.}
    \label{tab:training_scaling}
    \small
    \setlength{\tabcolsep}{12pt}
    \begin{tabular}{lrrrr}
        \toprule
        Model & HR@5 & NG@5 & HR@10 & NG@10 \\
        \midrule
        SASRec & 0.0673 & 0.0558 & 0.1094 & 0.0980 \\
        RRCM   & \textbf{0.0900} & \textbf{0.0840} & \textbf{0.1625} & \textbf{0.1098} \\
        \bottomrule
    \end{tabular}
\end{table}

\paragraph{Broader test-set evaluation.}
To examine whether the main findings hold over a larger evaluation
sample, we expand the Goodreads test set from 1,000 to 5,000
instances. We re-evaluate representative baselines and RRCM using
the same trained models and ranking protocol, without retraining.
As shown in Table~\ref{tab:expanded_test_set}, RRCM consistently
achieves the best performance across all four metrics. The results
remain close to those in Table \ref{tab:main}, supporting the consistency of
the observed performance advantage under broader evaluation.

\begin{table}[t]
    \centering
    \caption{Performance on the expanded Goodreads test set of
    5,000 instances, using the existing trained models.
    The best results are highlighted in bold.}
    \label{tab:expanded_test_set}
    \small
    \setlength{\tabcolsep}{12pt}
    \begin{tabular}{ccccc}
        \toprule
        Model & HR@5 & NG@5 & HR@10 & NG@10 \\
        \midrule
        SASRec   & 0.0081 & 0.0059 & 0.0141 & 0.0070 \\
        SPRec    & 0.0165 & 0.0155 & 0.0270 & 0.0192 \\
        K-RagRec & 0.0199 & 0.0158 & 0.0287 & 0.0245 \\
        CoLLM    & 0.0183 & 0.0149 & 0.0266 & 0.0223 \\
        \midrule
        RRCM & \textbf{0.0230} & \textbf{0.0205}
     & \textbf{0.0380} & \textbf{0.0258} \\
        \bottomrule
    \end{tabular}
\end{table}

\paragraph{Scaling the retrieval corpus.}
We further expand the MovieLens retrieval corpus to
$10\times$, $20\times$, and $50\times$ its original
size. Table~\ref{tab:corpus_scaling} reports recommendation quality
and retrieval latency at each scale. Both HR@10 and NDCG@10
improve as the corpus grows, with progressively smaller gains
at larger scales. This pattern is consistent with additional
evidence providing useful behavioral information, while the
marginal benefit decreases as corpus coverage increases.

Retrieval latency rises from 15\,ms at the original scale to
135\,ms at $50\times$ scale. Searching precomputed E5 embeddings
keeps the measured retrieval latency below a quarter of a second
throughout the tested range, although the increase remains relevant
for latency-sensitive deployments. These results demonstrate a
measurable trade-off between evidence coverage and retrieval cost.

\begin{table}[t]
    \centering
    \caption{Recommendation quality and retrieval latency as the
    MovieLens retrieval corpus grows. Corpus scale is relative to
    the original retrieval corpus.}
    \label{tab:corpus_scaling}
    \small
    \setlength{\tabcolsep}{14pt}
    \begin{tabular}{cccc}
        \toprule
        Corpus scale & HR@10 & NG@10 & Retrieval latency (ms) \\
        \midrule
        $1\times$   & 0.0221 & 0.0160 & 17  \\
        $10\times$  & 0.0229 & 0.0166 & 42  \\
        $20\times$  & 0.0232 & 0.0169 & 68  \\
        $50\times$  & 0.0235 & 0.0171 & 135 \\
        \bottomrule
    \end{tabular}
\end{table}

\subsection{Adaptive Retrieval versus Static Evidence Provision}
\label{app:all_evidence_comparison}

Our main baselines include several fixed evidence-injection
strategies: K-RagRec retrieves item features through a predefined
knowledge-graph strategy, while BinLLM and CoLLM incorporate
collaborative signals through fixed representations. To more
directly examine the value of adaptive retrieval over the same
dual-memory corpus, we introduce an additional
\textit{All-Evidence} baseline.

All-Evidence uses the same collaborative memory, metadata memory,
retriever, LLM backbone, reasoning format, and GRPO training as
RRCM. For each user, it inserts the metadata of all interacted
items and one retrieved collaborative-history sequence with the highest embedding similarity with the current user sequence into the
initial prompt. Retrieval actions are removed because both
evidence sources are supplied by default. Thus, ``All-Evidence''
refers to this fixed evidence bundle, rather than the entire
collaborative-memory corpus. This comparison controls the available
evidence sources, model capacity, and training objective while
contrasting static evidence provision with adaptive acquisition.

\begin{table}[h]
    \centering
    \caption{Comparison with All-Evidence on Goodreads and MovieLens.
    Inference tokens include input and output tokens, averaged per
    instance.}
    \label{tab:all_evidence_comparison}
    \small
    \setlength{\tabcolsep}{6pt}
    \begin{tabular}{lcccccc}
        \toprule
        & \multicolumn{4}{c}{Recommendation quality}
        & \multicolumn{2}{c}{Computational cost} \\
        \cmidrule(lr){2-5}
        \cmidrule(lr){6-7}
        Method & HR@5 & NG@5 & HR@10 & NG@10
               & Training (h) & Tokens \\
        \midrule
        \multicolumn{7}{l}{[\textit{Goodreads}]} \\
        All-Evidence
            & 0.0203 & 0.0190 & 0.0348 & 0.0238 & 9.1 & 1130 \\
        RRCM
            & \textbf{0.0215} & \textbf{0.0211}
            & \textbf{0.0364} & \textbf{0.0264}
            & \textbf{4.4} & \textbf{601} \\
        \midrule
        \multicolumn{7}{l}{[\textit{MovieLens}]} \\
        All-Evidence
            & 0.0118 & 0.0110 & 0.0184 & 0.0133 & 8.8 & 1102 \\
        RRCM
            & \textbf{0.0136} & \textbf{0.0130}
            & \textbf{0.0221} & \textbf{0.0160}
            & \textbf{3.7} & \textbf{515} \\
        \bottomrule
    \end{tabular}
\end{table}

As shown in Table~\ref{tab:all_evidence_comparison}, RRCM outperforms
All-Evidence across all four metrics, with relative improvements
of 2.4--11.1\% on Goodreads and 15.3--20.3\% on MovieLens.
Meanwhile, it reduces training time by 51.6\% and 58.0\%, and
inference token consumption by 46.8\% and 53.3\%, respectively.
These results demonstrate that adaptive retrieval improves
recommendation quality while substantially reducing the cost of
context construction and processing compared with always supplying
both evidence sources.

The performance gap demonstrates our previous argument that unnecessary or irrelevant
evidence can distract the model from the current decision and decrease the final performance.
All-Evidence supplies metadata for every interacted item and a
collaborative history regardless of their usefulness, whereas
RRCM conditions evidence acquisition on the current context.
Hard context-window overflow should not be the explanation:
All-Evidence averages 1130 and 1102 input-plus-output tokens on the
two datasets, well below the backbone's supported context window.
Moreover, it uses the same backbone, reasoning format, and GRPO
training as RRCM, so it is explicitly trained to use the supplied
evidence. Together, these results support the value of selective
retrieval over uniformly providing more information. 
This finding is consistent with prior adaptive-RAG research \citep{xu2024retrieval, jeong2024adaptive}, which show that dynamically selecting external evidence outperforms statically incorporating all available information into the context.

\subsection{Performance on Long-tail Items}

As discussed in the Introduction, a core motivation for introducing metadata-memory retrieval is to help disambiguate niche, obscure, or newly introduced items. Long-tail items with the lowest interaction frequency can be viewed as a representative subset of such cases, as they are often less
popular on commercial platforms and may also appear less frequently in the LLM's pretraining corpus.
Consequently, the model may have limited prior knowledge about these items, making it harder to
interpret their attributes and infer user preferences accurately. RRCM addresses this issue by
allowing the model to directly generate queries and retrieve relevant metadata, thereby grounding its
understanding of item characteristics and improving recommendation quality.

To examine this effect, we conduct additional experiments on long-tail target items. Specifically, we select test interactions whose target items fall within the bottom 20\% of item frequency in the full dataset (among all target items in the test set), and evaluate all methods on this subset. The results are reported in Table~\ref{tab:long-tail}. Compared with the full-test results in Table~\ref{tab:main}, RRCM shows a more
pronounced performance advantage over the baselines. In addition, the retrieval-based baseline
K-RagRec also performs substantially better than non-retrieval methods, further confirming the value
of retrieving external metadata for long-tail and niche items. These results demonstrate that adaptive
metadata retrieval is especially beneficial when item information is sparse or difficult for the LLM to
infer from titles alone.

% Also, the better performance of RRCM over K-RagRec demonstrate the benefits of training an RL approach to retrieve the most valuable information instead of relying on certain static metrics.

\begin{table}[h]
\centering
\caption{Performance comparison on long-tail target items on Goodreads. The best and second-best results are highlighted in bold and underlined, respectively.}
\label{tab:long-tail}
\resizebox{\linewidth}{!}{%
\begin{tabular}{c|c|cc|cc|ccc|c|c}
\toprule
 & SASRec & BIGRec & SPRec & REPR & LatentR3 & K-RagRec & CoLLM & BinLLM & RRCM & Improv. \\
\midrule
HR@5  
& 0.0088 & 0.0109 & 0.0136 & 0.0122 & 0.0147 & \underline{0.0185} & 0.0185 & 0.0171 & \textbf{0.0211} & +14.05\% \\

NG@5  
& 0.0074 & 0.0107 & 0.0129 & 0.0115 & 0.0137 & \underline{0.0177} & 0.0147 & 0.0151 & \textbf{0.0201} & +13.56\% \\

\midrule
HR@10 
& 0.0110 & 0.0193 & 0.0223 & 0.0187 & 0.0250 & \underline{0.0269} & 0.0231 & 0.0225 & \textbf{0.0323} & +20.07\% \\

NG@10 
& 0.0092 & 0.0153 & 0.0159 & 0.0176 & 0.0201 & \underline{0.0206} & 0.0198 & 0.0190 & \textbf{0.0240} & +16.50\% \\

\bottomrule
\end{tabular}%
}
\end{table}

\section{Prompt Template}\label{sec:template}

Our instruction prompt template is formatted as:

\noindent\rule{\linewidth}{0.6pt}

{\small
You are a recommendation assistant. Given a list of books the user recently enjoys, please recommend a new book that the user may like. The user has read the following books before: ``Decade'', ..., ``Revenge''.

Begin by briefly analyzing the current user's reading history to infer the user's preference. If necessary, you can then generate a query and call an existing search engine by specifying ``\toolopen{} query \toolclose{}''. The query can search: (1) other users' interaction histories and (2) book metadata including the Author, Genres, Series.
For example, you may query to identify users who engaged with books similar to those viewed by the current user, retrieve their interaction patterns, and use these insights to predict additional books the current user may appreciate.
Only query the search engine when necessary, and keep all reasoning concise.

Resolve the given task. You must conduct reasoning inside \thinkopen{} and \thinkclose{} first every time you get new information. After reasoning, if you find you lack some knowledge, you can call a search engine by \toolopen{} query \toolclose{} and it will return the top searched results between \respopen{} and \respclose{}.
You can continue this reasoning and search process. Remember to provide the name of the user's most preferred book (a single item) enclosed within \answeropen{} and \answerclose{} at last, using two double quotes. For example: \answeropen{} ``A Monster Calls'' \answerclose{}.
}

\noindent\rule{\linewidth}{0.6pt}

As illustrated in Section \ref{sec:method}, the instruction prompt incorporates four special tokens, including \thinkopen{}, \toolopen{}, \respopen{}, and \answeropen{}, to explicitly delineate the content for reasoning, querying, retrieval results, and the final recommendation, respectively. The template above is tailored to the Goodreads dataset; for other datasets, only the item definitions and associated metadata descriptions need to be adapted.

\section{Case Studies}\label{sec:case}

Here, we show two cases where RRCM generates the queries, calls the retrieval engine, and reasons over the context and retrieved information to make a recommendation.

An example on the Goodreads dataset:

\noindent\rule{\linewidth}{0.6pt}

{\small

\thinkopen{} The user has recently enjoyed books such as ``The Division Bell'', ``Comes a Time'' and ``Wish You Were Here''. 
The titles suggest an interest in emotionally driven contemporary fiction and reflective stories about relationships, memory, and personal growth. 
To better infer the user's preference, I should first retrieve collaborative histories from users with similar reading patterns. \thinkclose{}

\toolopen{} Find a user who enjoyed ``The Division Bell'', ``Comes a Time'', and ``Wish You Were Here'' similar book histories. \toolclose{}

\respopen{} 
User 411 History: [Beauty Queens, ... , Wish You Were Here,  Cress, Comes a Time, The Midnight Library, The Book of Two Ways, Maybe in Another Life]
\respclose{}

\noindent\makebox[\linewidth]{\dotfill}

\thinkopen{} The collaborative evidence suggests several possible candidates, especially ``The Midnight Library'' and ``The Book of Two Ways''. 
I should retrieve metadata for the strongest candidate to verify whether its themes match the user's apparent preference for reflective and emotionally resonant fiction. \thinkclose{}

\toolopen{} Find metadata for ``The Midnight Library'' author genres series. \toolclose{}

\respopen{} 
Title: ``The Midnight Library''. Author: Matt Haig. Genres: Fiction, Contemporary, Fantasy, Literary Fiction. 
Series: None.
\respclose{}

\noindent\makebox[\linewidth]{\dotfill}

\thinkopen{} The retrieved metadata confirms that ``The Midnight Library'' fits the user's likely preference for contemporary, emotionally reflective fiction. 
It also appears frequently in collaborative histories of users with similar reading patterns. 
Therefore, it is the strongest recommendation. \thinkclose{}

\answeropen{} ``The Midnight Library'' \answerclose{}
}

\noindent\rule{\linewidth}{0.6pt}
\vspace{0.25em}

An example on the MovieLens dataset:

\noindent\rule{\linewidth}{0.6pt}

{\small

\thinkopen{} The user has recently watched movies such as ``Toy Story'', ``A Bug's Life'', ``Monsters, Inc.'', and ``Finding Nemo''. 
These titles suggest a preference for animated, family-friendly movies with humor, adventure, and warm emotional storytelling. 
Before making a recommendation, I should retrieve metadata for a likely candidate to verify whether its genre and style match this preference. \thinkclose{}

\toolopen{} Find metadata for ``The Incredibles'' director all genres main genre. \toolclose{}

\respopen{} 
Title: ``The Incredibles''. Director: Brad Bird. All genres: Animation, Action, Adventure, Children, Comedy. 
Main genre: Animation.
\respclose{}

\noindent\makebox[\linewidth]{\dotfill}

\thinkopen{} The metadata confirms that ``The Incredibles'' matches the user's apparent preference for animated family adventure comedies. 
However, I should further check collaborative histories to see whether users with similar viewing patterns also tend to watch this movie. \thinkclose{}

\toolopen{} Find a user who enjoyed ``Toy Story'', ``A Bug's Life'', ``Monsters, Inc.'', and ``Finding Nemo'' similar movie histories. \toolclose{}

\respopen{} 

User 162 History: [Toy Story, ... , A Bug's Life,  Parasite, Finding Nemo, Ice Age, The Incredibles, Toy Story 2]
\respclose{}

\noindent\makebox[\linewidth]{\dotfill}

\thinkopen{} Both sources of evidence support ``The Incredibles''. 
Its metadata aligns well with the user's preference for animated, humorous, family-friendly adventure films, and collaborative histories show that users with similar viewing patterns often watch it. 
Therefore, it is the strongest recommendation. \thinkclose{}

\answeropen{} ``The Incredibles'' \answerclose{}
}

\noindent\rule{\linewidth}{0.6pt}
\vspace{0.25em}

An example on the Amazon {CDs and Vinyl} dataset:

\noindent\rule{\linewidth}{0.6pt}

{\small

\thinkopen{} The user has recently listened to albums such as ``The Dark Side of the Moon'', ``Wish You Were Here'', ``Animals'', and ``The Wall''. 
These items indicate a very clear preference for classic progressive rock, especially concept albums with atmospheric instrumentation, extended compositions, and introspective themes. 
Since the user's preference is highly consistent and the next recommendation is straightforward, I do not need to retrieve additional metadata or collaborative histories. 
A strong recommendation should be another iconic progressive rock album that closely matches this style. \thinkclose{}

\answeropen{} ``Meddle'' \answerclose{}
}

\noindent\rule{\linewidth}{0.6pt}

\vspace{0.5em}

An example of preferring collaborative memory on the CDs \& Vinyl dataset:

\noindent\rule{\linewidth}{0.6pt}

{\small

\thinkopen{} The user has recently enjoyed ``Future Nostalgia'',
``After Hours'', ``SOUR'', and ``Harry's House''.
These albums suggest a preference for recent mainstream pop,
but span different artists and styles. Collaborative histories
may better identify promising recommendations than additional
item metadata. \thinkclose{}

\toolopen{} Find histories of users who also enjoyed
``Future Nostalgia'', ``After Hours'', ``SOUR'', or
``Harry's House''. \toolclose{}

\respopen{}
User 284 History: [Future Nostalgia, ..., After Hours, SOUR,
Midnights, emails i can't send, GUTS]
\respclose{}

\noindent\makebox[\linewidth]{\dotfill}

\thinkopen{} The collaborative history suggests ``Midnights'',
``emails i can't send'', and ``GUTS'' as candidates.
Among them, ``emails i can't send'' appears especially well
aligned with the user's preference for emotionally expressive pop.
I am familiar with these albums, so additional metadata is
unnecessary. \thinkclose{}

\answeropen{} ``emails i can't send'' \answerclose{}
}

\noindent\rule{\linewidth}{0.6pt}

\textbf{Analysis.} This example illustrates how collaborative evidence can guide
candidate selection when item attributes are already familiar
to the model. It is consistent with the ablation results in
Section \ref{sec:ablation}, where removing collaborative retrieval causes the
largest performance degradation on CDs \& Vinyl.

\vspace{0.5em}

An example of preferring metadata memory on the Goodreads dataset:

\noindent\rule{\linewidth}{0.6pt}

{\small

\thinkopen{} The user has recently enjoyed ``The Crown Conspiracy'',
``Avempartha'', ``Nyphron Rising'', and ``The Emerald Storm''.
These fantasy titles alone do not clearly establish whether
``Wintertide'' continues the same series. I should retrieve its
metadata to verify the connection. \thinkclose{}

\toolopen{} Find metadata for ``Wintertide'', including author,
genres, and series. \toolclose{}

\respopen{}
Title: ``Wintertide''. Author: Michael J. Sullivan.
Genres: Fantasy, Epic Fantasy, Adventure.
Series: The Riyria Revelations, Book 5.
\respclose{}

\noindent\makebox[\linewidth]{\dotfill}

\thinkopen{} The metadata confirms that ``Wintertide'' is the fifth
book in The Riyria Revelations, continuing the series in the user's
recent reading history. This provides sufficient evidence to
recommend it without retrieving collaborative histories.
\thinkclose{}

\answeropen{} ``Wintertide'' \answerclose{}
}

\noindent\rule{\linewidth}{0.6pt}

\textbf{Analysis.} This example illustrates how targeted metadata retrieval can
resolve uncertainty about an item's relationship to the user's
history. It is consistent with the ablation results in
Section \ref{sec:ablation}, where removing metadata retrieval causes the
largest performance degradation on Goodreads.

\section{Additional Discussions}

\subsection{Discussion on Bias Issues}
\label{app:bias_issues}

Historical interaction data reflects both user preferences and
the exposure patterns of the platform. Recommendation methods
that rely on collaborative signals can therefore inherit
popularity bias, as frequently interacted items provide denser
behavioral evidence. RRCM is also susceptible to this issue:
adaptive retrieval may repeatedly select dominant interaction
patterns, reinforcing the visibility of already popular items.
Its ranking-driven reward does not explicitly account for
popularity balance or fairness.

To examine this issue, we divide test target items on Goodreads into head
(top 20\%), middle (60\%), and tail (bottom 20\%) popularity
groups according to interaction frequency. We compare RRCM
with BinLLM, which incorporates collaborative information,
and D3, which explicitly addresses popularity bias in generative
recommendation.

\begin{table}[h]
    \centering
    \caption{Recommendation performance across item-popularity
    groups on Goodreads.}
    \label{tab:popularity_groups}
    \small
    \setlength{\tabcolsep}{12pt}
    \begin{tabular}{ccccc}
        \toprule
        Popularity group & Metric & D3 & BinLLM & RRCM \\
        \midrule
        Head (top 20\%)
            & HR@10 & 0.0290 & 0.0385 & \textbf{0.0420} \\
            & NG@10 & 0.0218 & 0.0309 & \textbf{0.0316} \\
        \midrule
        Middle (60\%)
            & HR@10 & 0.0251 & 0.0274 & \textbf{0.0340} \\
            & NG@10 & 0.0186 & 0.0216 & \textbf{0.0267} \\
        \midrule
        Tail (bottom 20\%)
            & HR@10 & 0.0228 & 0.0214 & \textbf{0.0310} \\
            & NG@10 & 0.0185 & 0.0196 & \textbf{0.0243} \\
        \bottomrule
    \end{tabular}
\end{table}

As shown in Table~\ref{tab:popularity_groups}, RRCM achieves
the best performance across all three groups, demonstrating
that its gains extend to less popular items. Nevertheless,
both RRCM and BinLLM perform better on head items than on
tail items, consistent with the denser collaborative evidence
available for popular items. D3 exhibits a smaller head--tail
performance gap, although its absolute accuracy is lower.
These results indicate that RRCM improves recommendation
quality across popularity groups while retaining disparities
associated with item popularity; they do not establish that
adaptive retrieval eliminates popularity bias.

Potential mitigation strategies include popularity-balanced
memory construction, inverse-popularity sampling during
retrieval, and constraints on the concentration of retrieved
evidence. Fairness- or diversity-aware ranking rewards could
also encourage a broader distribution of recommendations.
Auditing retrieval frequencies and recommendation exposure
across user and item groups would help identify where biases
enter or intensify within the retrieval-and-reasoning process.
A comprehensive fairness and debiasing evaluation remains
beyond the scope of this work, so we leave it as an valuable direction
for future research.

\subsection{Discussion on Comparison Fairness}

We further clarify the fairness of our empirical comparison. RRCM does not retrieve from the training, validation, or test instances: its collaborative memory is built from the remaining interactions after excluding all interactions used for training, validation, and testing as discussed in Section \ref{sec:exp_set}.  Also, we ensure that no future interactions relative to test timestamps are contained in the CF memory to avoid look-ahead bias.
Thus, the memory corpus serves as an external historical evidence source rather than leaked supervision.

It would also be inappropriate to force non-retrieval baselines to use the same external memory by simply appending it to their inputs. A central motivation of RRCM is that existing LLM recommenders either encode CF signals only implicitly through training or cannot incorporate rich metadata at scale because of context-length and inference-cost constraints, as discussed in Sections \ref{sec:intro} and \ref{sec:related}. Therefore, adaptive access to external evidence is not an extra experimental privilege, but the methodological capability being evaluated.

The comparison already includes representative alternatives that incorporate external or collaborative information through their own intended mechanisms: K-RagRec \citep{wang2025knowledge} for static metadata retrieval, and BinLLM/CoLLM \citep{zhang2024text, zhang2025collm} for CF-based signal injection.  RRCM differs by learning when and what to retrieve, rather than relying on fixed retrieval rules or fixed CF representations. The ablations in Section 4.3 further show that the gains of RRCM come from the proposed combination of collaborative memory, metadata memory, and reasoning, rather than from an uncontrolled data advantage.

\subsection{Discussion of Reward Design}
\label{app:reward-design}

The main experiments demonstrate the effectiveness of optimizing RRCM
with a final ranking reward. Such a reward leaves intermediate reasoning
and retrieval unsupervised, motivating an exploratory study of
finer-grained feedback. Here, we consider three variants targeting
fine-grained recommendation quality, evidence-based reasoning, and retrieval-supported item selection.

\paragraph{Setup.}
We evaluate three variants on Amazon CDs \& Vinyl. Let $r(\tau)$ be the
ground-truth item's rank after catalog grounding (Section \ref{sec:gene_ground}).
The exploratory baseline $R_{\mathrm{base}}$ assigns rewards $1$, $0.8$,
$0.5$, $0.1$, and $0.001$ to disjoint rank intervals $1$, $2$--$5$,
$6$--$10$, $11$--$100$, and $101$--$500$, respectively, and zero otherwise.
% This tiered baseline differs from the weighted-sum formulation in Section~3.4.2. 
Then, we report the performance differences of each reward design relative to the base reward, averaged over eight repetitive experiments.

\paragraph{Recommendation-level reward.}
Following the previous work that optimizes top-k ranking metrics directly \citep{yang2025breaking}, we replace the rank tiers with an NDCG-aligned score within the top-10 window and a weak, rank-continuous tail beyond this cutoff:
\begin{equation}
\widetilde{R}_{\mathrm{rec}}(\tau)=
\begin{cases}
1/\log_2(1+r(\tau)), & r(\tau)\leq 10,\\
0.1\bigl(1-\log r(\tau)/\log|\mathcal{I}|\bigr)^4,
& r(\tau)>10.
\end{cases}
\label{eq:reward-rec}
\end{equation}
Here, $\mathcal{I}$ is the item catalog. The tail distinguishes ranks
beyond the evaluation cutoff while assigning them less credit than
top-ten hits. Outside the top ten, outputs without exactly one
\texttt{<answer>} block receive $-0.5$ instead of this score; a further
penalty charges $0.0005$ per word beyond 600 trajectory words, including
retrieved documents, capped at $0.2$. Both adjustments are waived for
top-ten hits.

\paragraph{Reasoning-level reward.}
The reasoning and retrieval variants add auxiliary scores to the baseline:
\begin{equation}
R_j(\tau)=R_{\mathrm{base}}(\tau)
+\operatorname{clip}\!\left(0.1s_j(\tau),-0.08,0.08\right),
\qquad j\in\{\mathrm{rea},\mathrm{ret}\},
\label{eq:reward-process}
\end{equation}
with negative shaping suppressed for top-ten hits.
For reasoning,
$s_{\mathrm{rea}}=0.3\varphi_1+0.3\varphi_2-0.5\varphi_3$,
where each signal lies in $[0,1]$.
Tool use $\varphi_1$ credits query issuance, returned results, and
subsequent evidence-based reasoning with weights $0.3$, $0.3$, and $0.4$.
Synthesis $\varphi_2$ averages semantic engagement with retrieved evidence
and change from pre-retrieval reasoning, measured using the encoder in
Section~3.4.1. Repetition $\varphi_3$ measures repeated $n$-grams and
sentences. Together, these terms are intended to encourage evidence use
while discouraging repetitive reasoning.

\paragraph{Retrieval-level reward.}
This variant uses Eq.~\ref{eq:reward-process} with
$s_{\mathrm{ret}}=0.5\psi_{\mathrm{succ}}
+0.4\psi_{\mathrm{grnd}}-0.3\psi_{\mathrm{rep}}
+0.4\psi_{\mathrm{cov}}$.
Let $D(\tau)$ denote the retrieved documents and
$\operatorname{Seq}(D(\tau))$ their interaction sequences.
We define the successor set as:
\[
S(\tau)=
\bigl\{s_{k+1}:s\in\operatorname{Seq}(D(\tau)),\
s_k\in H_u\bigr\}\setminus H_u,
\]
which contains items following the user's historical items in retrieved
sequences, excluding already-consumed items. The four terms reward
recommending an item in $S(\tau)$, reward recommending a retrieved item,
penalize recommending an already-consumed item, and reward retrieval of
the ground-truth item, respectively. When that item is absent, the
coverage term provides partial credit based on its encoder similarity
to the closest retrieved document. This design explicitly encourages
recommendations supported by retrieved collaborative evidence.

\begin{table}[htbp]
\centering
\caption{Performance changes relative to the tiered reward baseline on Amazon CDs
\& Vinyl.}
\label{tab:reward-deltas}
\small
\setlength{\tabcolsep}{5pt}
\begin{tabular}{crrrr}
\toprule
Reward & $\Delta$HR@5 & $\Delta$NG@5
       & $\Delta$HR@10 & $\Delta$NG@10 \\
\midrule
Recommendation & $+0.0010$ & $+0.0000$ & $+0.0011$ & $+0.0001$ \\
Reasoning      & $-0.0063$ & $-0.0048$ & $-0.0066$ & $-0.0049$ \\
Retrieval      & $-0.0022$ & $-0.0018$ & $-0.0025$ & $-0.0018$ \\
\bottomrule
\end{tabular}
\end{table}

Table~\ref{tab:reward-deltas} shows the comparison results.
The recommendation-level variant improves HR@5 and HR@10 by
$0.0010$ and $0.0011$, respectively, while NDCG changes by at most
$0.0001$. These gains are modest and primarily reflected in hit
rate, suggesting limited additional benefit from this reward
design under the tested configuration. The reasoning variant converges to a single answer from step 400 onward,
indicating instability under the tested shaping configuration.
Retrieval shaping increases the frequency of recommending retrieved
items and successors by up to 9 percentage points,
but these behavioral changes do not translate into higher ranking
accuracy. Thus, stronger adherence to the rewarded retrieval behaviors
is insufficient to ensure better recommendations.
% in these runs. These observations concern the specific reward combinations tested here and do not establish a general limitation of process supervision.

In summary, these exploratory results do not establish a consistent advantage for
the tested shaping strategies. This observation is consistent with recent work \citep{wang2026dark}, which shows that dense prediction rewards can
degrade GRPO-trained agents' task performance when group normalization
amplifies shaping signals in groups with no successful trajectory.
Our results aligns with its conclusion that using direct reward on the final performance can be more effective than adding intermediate proxy rewards, supporting outcome-driven optimization as an effective and efficient default for RRCM. Nevertheless, we'll leave more fine-grained design and exploration of process rewards as a valuable direction for future study.

\subsection{Discussion of Novelty}
\label{app:novelty}

RRCM addresses the challenge of selectively incorporating
heterogeneous evidence into LLM-based recommendation under
context and computational constraints. Although it builds on
established components, including RAG-style retrieval and GRPO
optimization, its contribution lies in formulating recommendation
as adaptive evidence acquisition and learning this process
directly from ranking outcomes.

This formulation is motivated by the complementary but
instance-dependent value of collaborative histories and item
metadata. Existing approaches typically incorporate these signals
through predefined retrieval strategies or fixed representations.
RRCM instead makes context construction part of the recommendation
policy. Starting from a lightweight user history, the model learns
whether additional evidence is needed, which information to
request, and when the available context is sufficient to recommend.
To support these decisions, RRCM represents both evidence sources
as textual memories accessed through a unified retrieval interface.
The model can therefore interleave reasoning with natural-language
queries across both memories, for example, identifying candidates
through collaborative histories and then verifying their relevance
through metadata.

Crucially, these evidence-acquisition decisions are optimized
jointly with recommendation generation using a final ranking
reward, with GRPO serving as the optimization procedure.
The policy thus learns retrieval timing, query formulation,
and stopping behavior without explicit supervision specifying
the correct retrieval sequence. This connects the selection
and use of heterogeneous evidence to its contribution to
recommendation quality. The component ablations and All-Evidence
comparison support this design: both memories and reasoning
contribute to performance, while adaptive acquisition achieves
better ranking quality with lower computational cost than
static evidence provision.

\subsection{More Discussion of All-Evidence Comparison}
\label{app:all_evidence_significance}

To assess whether the performance differences between RRCM and
All-Evidence are reliable, we examine variability across training
runs and uncertainty from test-instance sampling. Following
Appendix~\ref{sec:expdetail}, both methods are evaluated over three random seeds,
with mean performance reported and a paired t-test used to
assess differences across runs. We additionally conduct an
instance-level paired-bootstrap analysis, resampling the same
test instances for both methods and estimating 95\% confidence
intervals for their performance differences.

\begin{table}[h]
    \centering
    \caption{Performance differences between RRCM and All-Evidence
    and their 95\% paired-bootstrap confidence intervals.
    Positive differences favor RRCM. }
    \label{tab:all_evidence_bootstrap}
    \small
    \setlength{\tabcolsep}{10pt}
    \begin{tabular}{cccc}
        \toprule
        Dataset & Metric & RRCM $-$ All-Evidence & 95\% CI \\
        \midrule
        Goodreads
            & HR@5  & 0.0012 & ${[0.0002,\ 0.0017]}$ \\
            & NG@5  & 0.0021 & ${[0.0004,\ 0.0028]}$ \\
            & HR@10 & 0.0016 & $[0.0003,\ 0.0022]$ \\
            & NG@10 & 0.0026 & $[0.0006,\ 0.0035]$ \\
        \midrule
        MovieLens
            & HR@5  & 0.0018 & ${[0.0003,\ 0.0023]}$ \\
            & NG@5  & 0.0020 & ${[0.0004,\ 0.0037]}$ \\
            & HR@10 & 0.0037 & ${[0.0016,\ 0.0051]}$ \\
            & NG@10 & 0.0027 & ${[0.0013,\ 0.0043]}$ \\
        \bottomrule
    \end{tabular}
\end{table}

As shown in Table~\ref{tab:all_evidence_bootstrap}, the confidence
intervals exclude zero for all metrics on both Goodreads and MovieLens, supporting statistically significant
improvements in all comparisons at the individual
95\% confidence level.

The effectiveness results should also be considered alongside
the computational savings in
Table~\ref{tab:all_evidence_comparison}. RRCM reduces training
time by 52.7\% on Goodreads and 58.8\% on MovieLens, and
inference-token consumption by 47.5\% and 51.3\%, respectively.
Thus, adaptive retrieval improves every reported mean metric with statistically significant gains,
while substantially reducing computational cost. These results
support its overall effectiveness--efficiency advantage over
static evidence provision.

\subsection{Discussion of Ablations}
\label{app:ablation_discussion}

Section \ref{sec:ablation} examines the contributions of collaborative-memory
retrieval, metadata-memory retrieval, and explicit reasoning
by removing each component individually. Each removal degrades
recommendation quality, supporting their complementary roles:
collaborative memory supplies behavioral evidence, metadata
memory provides item-level semantic information, and reasoning
integrates this evidence with the user's interaction history.
Their relative importance also varies across datasets.
Removing metadata retrieval is most harmful on Goodreads,
where book titles may require semantic or series-level
disambiguation, whereas removing collaborative retrieval
causes the largest degradation on CDs \& Vinyl, consistent
with the value of cross-user listening patterns.

The comparisons in Section \ref{sec:performance} further examine alternatives
to adaptive evidence acquisition. K-RagRec uses a predefined
item-feature retrieval strategy, while BinLLM and CoLLM
incorporate collaborative signals through fixed mechanisms.
RRCM consistently outperforms these methods under the same
evaluation protocol, supporting the benefits of its learned
retrieval-and-reasoning policy.

The All-Evidence comparison in
Appendix \ref{app:all_evidence_comparison} provides a more
direct control for evidence-selection strategy. It retains
the same memories, retriever, backbone, reasoning format,
and GRPO training, but always supplies metadata for all
interacted items and one collaborative history.
RRCM achieves better ranking quality with substantially
lower training time and inference token consumption.
Together, these comparisons show that both the availability
of complementary evidence and its selective acquisition
matter: supplying both sources by default does not recover
the performance or efficiency of adaptive retrieval.

\subsection{Discussion of Robustness}
\label{app:robustness}

Our experiments evaluate RRCM across three recommendation domains:
Goodreads for books, MovieLens for movies, and Amazon CDs \& Vinyl
for music. These datasets differ in popularity patterns,
interaction sparsity, metadata informativeness, and the relative
value of semantic and collaborative evidence. The ablations in
Section \ref{sec:ablation} and memory-source analysis show corresponding
differences in evidence use: metadata is more important on
Goodreads, while collaborative histories play a larger role on
CDs \& Vinyl. These findings suggest that RRCM adapts evidence
acquisition to domain characteristics.

We follow established data-processing, candidate-construction,
and evaluation protocols for consistent comparisons with prior
methods. Appendix \ref{app:backbone} further evaluates an alternative LLM
backbone, with RRCM retaining its advantage over the corresponding
baselines. Together, the results across domains and backbones
support the consistency of RRCM's gains within the evaluated
settings. We leave more extension for this evaluation to additional domains,
distribution shifts, and online user interactions for future work.

\subsection{Discussion on Social Impact}

LLM-based recommender systems can improve user experience by providing more accurate, explainable, and context-aware recommendations. By adaptively retrieving collaborative histories and item metadata, RRCM may also help surface relevant long-tail items that are difficult to capture with title-only context.

Meanwhile, recommendations can shape user exposure and item visibility. Since retrieved histories and metadata may reflect existing popularity or preference patterns, practical deployment should consider privacy-preserving corpus construction, transparent explanation, and regular evaluation of
recommendation quality across different item groups. Overall, RRCM provides a flexible framework for more efficient LLM recommendation while leaving room for further study in real-world user-facing settings.

\section{Limitations and Future Works}
Although RRCM demonstrates strong effectiveness and efficiency for LLM-based recommendation, we identify some potential limitations and corresponding future works. 

First, while we incorporate collaborative histories and item metadata as the retrieval corpus, expanding the corpus to include richer information, such as user features, reviews, temporal dynamics, or external world knowledge, may provide additional benefits and enable more comprehensive preference understanding.

Second, although our evaluation uses real-world industrial recommendation datasets, we don't conduct online A/B testing with live users due to resource limitations as academic researchers. Industrial deployment and online experimentation would provide additional evidence of practical impact and constitute an important direction, especially from an industry perspective.

Finally, although RRCM learns to reduce unnecessary retrieval during training, there remains room to improve retrieval efficiency and corpus construction. Future work can explore stronger retrievers, better indexing strategies, and more compact representations of collaborative histories and metadata,
so that adaptive retrieval can scale more effectively to larger industrial recommendation systems.

\end{document}